\documentclass[aps,prd,nofootinbib,showkeys,floatfix,twocolumn]{revtex4-1}
\usepackage{amsmath}
\usepackage{amssymb}
\usepackage{graphicx}
\usepackage{xspace}
\usepackage{bbm} 
\usepackage{color}
\usepackage[utf8]{inputenc}
\usepackage{cancel}
\usepackage{slashed}
\usepackage{soul}
\usepackage{siunitx}

\usepackage{tikz} 
\usetikzlibrary{chains,shapes,fit,calc}
\usetikzlibrary{positioning,arrows}

\usetikzlibrary{patterns,decorations.pathreplacing}
\usetikzlibrary{decorations.pathmorphing}
\usetikzlibrary{decorations.markings}
\usetikzlibrary{arrows,shapes}
\usetikzlibrary{shapes.misc}
\tikzset{
    gluon/.style={decorate, draw=black,
        decoration={coil,amplitude=4pt, segment length=4pt,aspect=0.7}} 
}
\tikzset{
    photon/.style={decorate, decoration={snake}},
}

\definecolor{mygray}{gray}{0.8}

\definecolor{mkgreen}{rgb}{0.2,.70,.3}

 \definecolor{fsblue}{rgb}{0.,.0,1.}

 \definecolor{morao}{rgb}{0.5,.0,1.}

\newcommand\SARAH{{\tt SARAH}\xspace}

\newcommand\SPheno{{\tt SPheno}\xspace}

\newcommand\checkmate{{\tt CheckMATE}\xspace}

\def\met{\slash\hspace*{-1.5ex}E_{T}}

\usepackage[pdftex,colorlinks,linkcolor=cyan,citecolor=blue,urlcolor=cyan]{hyperref} 
\hypersetup{linktocpage}
\usepackage[capitalise, english]{cleveref}
\crefname{section}{Sec.}{Secs.}
\crefname{table}{Tab.}{Tabs.}

\begin{document}
\vspace{1cm}

\title{\LARGE Illuminating {\textcolor{mygray}{Stealth}} Scenarios at the LHC}

\hfill \parbox{5cm}{\vspace{ -1cm } \flushright BONN-TH-2018-11 \\ KA-TP-39-2018}

\newcommand{\AddrBonn}{%
Bethe Center for Theoretical Physics \& Physikalisches Institut der 
Universit\"at Bonn,\\ Nu{\ss}allee 12, 53115 Bonn, Germany}

\newcommand{\AddrSA}{
National Institute for Theoretical Physics,\\
School of Physics,
University of the Witwatersrand, Johannesburg, Wits 2050, South Africa}

\newcommand{\AddrKAITP}{
Institute for Theoretical Physics (ITP), 
Karlsruhe Institute of Technology, \\
Engesserstra{\ss}e 7, D-76128 Karlsruhe, Germany}

\newcommand{\AddrKAIKP}{
Institute for Nuclear Physics (IKP), Karlsruhe Institute of Technology,\\
Hermann-von-Helmholtz-Platz 1, D-76344 Eggenstein-Leopoldshafen, Germany}

\author{Jong Soo Kim} \email{jongsoo.kim@tu-dortmund.de}
\affiliation{\AddrSA}%

\author{Manuel E. Krauss} \email{mkrauss@th.physik.uni-bonn.de} 
\affiliation{\AddrBonn}

\author{V\'ictor Mart\'in Lozano} \email{lozano@physik.uni-bonn.de} 
\affiliation{\AddrBonn}

\author{Florian Staub}\email{florian.staub@kit.edu}
\affiliation{\AddrKAITP}\affiliation{\AddrKAIKP}%

\begin{abstract}
Several ideas exist how the stringent mass limits from LHC on new coloured particles can be avoided. One idea are the so-called `stealth' scenarios in which missing transversal energy ($\met$) is avoided due a peculiar mass configuration. It is usually assumed that the cascade decay of the dominantly-produced coloured particle finishes in a two-body decay, where this mass configuration  leads to a very small amount of $\met$. We discuss here the potential impact of other decay channels, either loop-induced or via off-shell mediators. It is shown that those channels already become important even for moderate branching ratios of 10\%. Larger branching ratios in particular into a photon can completely wash out all benefits of the stealth setup. We discuss this in a model-independent form, but also at the simplest SUSY stealth scenario which can be realised in the NMSSM.
\end{abstract}

\maketitle


\section{Introduction}
The Large Hadron Collider (LHC) has now collected data since more than eight years. While the long searched-for Higgs boson of the standard model (SM) of particle physics has been discovered  after two years of runtime \cite{Aad:2012tfa,Chatrchyan:2012xdj}, no clear signal for new physics has shown up so far. This is surprising because there is overwhelming evidence that the SM must be extended, e.g. to explain dark matter or the baryon asymmetry in the Universe. Also the hierarchy problem is an unresolved question. Many ideas to address these problems predict the presence of additional scalars at -- or at least close to -- the electroweak (ew) scale. Therefore, having only null results in the searches for beyond-the-SM (BSM) physics was unexpected, and scenarios which were considered to be likely have been ruled out by now. The best example is minimal supersymmetry with moderately light masses: benchmark scenarios developed for the LHC like SPS1a used squark and gluino masses of 600\,GeV and below \cite{Allanach:2002nj}, while the exclusion limits of these particles have reached up to 2.0\,TeV under specific conditions \cite{CMS:2017kmd,Aaboud:2017vwy}. This has tremendous consequences and many well-studied scenarios become disfavoured as solutions for the open issues in the SM. In order not to give up the appealing aspects of these ideas, approaches were discussed how the strong exclusion limits could be avoided. Since many searches for new physics rely on large amounts of missing transversal energy ($\met$), a promising ansatz is to reduce it as much as possible.  $R$-parity violation, which opens decay channels of the lightest supersymmetric particle (LSP), reduces the mass limits at least to some extent \cite{Dercks:2017lfq,Hanussek:2012eh,Hanussek:2012fc}, but revives the problem of a missing DM candidate. On the other side, compressed spectra could also shrink the MET significantly \cite{Carena:2008mj,Bornhauser:2010mw,Drees:2012dd,Aaboud:2017phn}, and could be motivated by relic density requirements that can be easily satisfied in the stop--neutralino co-annihilation region \cite{Boehm:1999bj}. 
Therefore, one can study models in which the lightest SUSY particle is very light, i.e. it has a mass of only a few GeV, and $\met$ is significantly reduced by a very specific kinematic configuration: the second decay product of the next-to-lightest SUSY (NLSP) particle almost fills the mass gap between the NLSP and LSP completely. If this particle is not visible (or at least hard to search for) at a collider, one has all ingredients for a so-called `stealth' scenario \cite{Fan:2011yu,CMS:2012un}. It has been pointed out in Ref.~\cite{Ellwanger:2014hia} that one does not need to introduce additional particles or even a hidden sector to have such a setup. Also in the next-to-minimal supersymmetric standard model (NMSSM) one could arrange for the necessary mass configuration: the bino NLSP can decay invisibly into a singlino LSP and a singlet. \\
This is, of course, a very attractive idea to soften the mass limits on the gluino in SUSY models. Since the focus in literature was so far only on the two-body decay of the NSLP, we study in this work the impact of additional decay modes either via loops or off-shell mediators. As we will show, one needs to consider these decay channels in order to be sure that the `stealth' mechanism is really working properly. \\
This paper is organised as follows. We start in \cref{sec:model-ind} with a model-independent study of the impact of three-body or loop-induced decays on stealth scenarios. Afterwards, we show in \cref{sec:examples} two examples where these additional decay modes become important.  We summarise our results in \cref{sec:conclusion}.



\section{Model-independent analysis}
\label{sec:model-ind}

\begin{figure}
\begin{center}
\begin{tikzpicture}
\node (Glu) at (1,10.) {\large $\tilde{g}$};
\node (NLSP) at (1,4.) {\large NLSP};
\node (S) at (4,3.5) {\large S};
\node (LSP) at (1,0.5) {\large LSP};
\draw [->, line width=0.5mm] (NLSP) -- (LSP);
\draw [->, line width=0.5mm] (NLSP) -- (S);
\draw [->, line width=0.5mm] (Glu) -- (NLSP);
\draw [-, line width=0.75mm] (-1,0) -- (-1,6);
\draw [-, line width=0.75mm] (-1,6) -- (-0.75,6.25);
\draw [-, line width=0.75mm] (-0.75,6.25) -- (-1.25,6.5);
\draw [-, line width=0.75mm] (-1.25,6.5) -- (-0.75,6.75);
\draw [-, line width=0.75mm] (-0.75,6.75) -- (-1.25,7);
\draw [-, line width=0.75mm] (-1.25,7.00) -- (-0.75,7.25);
\draw [-, line width=0.75mm] (-0.75,7.25) -- (-1,7.5);
\draw [->, line width=0.75mm] (-1,7.5) -- (-1,11);
\draw [-, line width=0.5mm] (-1.5,0) -- (-0.5,0);
\draw [-, line width=0.5mm] (-1.5,4.5) -- (-0.5,4.5);
\draw [-, line width=0.5mm] (-1.5,9) -- (-0.5,9);
\node (LSP) at (-1.75,11) {\large $M$};
\node (LSP) at (-1.95,4.5) {\large $M_Z$};
\node (LSP) at (-2.25,9) {\large $M_{\rm SUSY}$};
\node (LSP) at (-1.75,0) {\large $0$};
\end{tikzpicture}
\end{center}
\caption{The kinematic configuration necessary for the stealth mechanism.}
\label{fig:massconfig}
\end{figure}
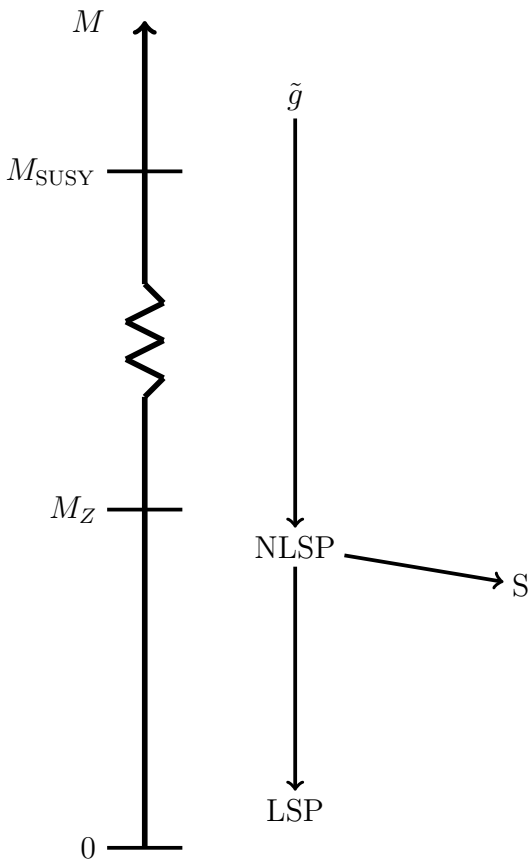

A typical stealth mass configuration is depicted in Fig.~\ref{fig:massconfig}. 
The mass scale of the SUSY particles, in particular the coloured ones,  is considered well above the $Z$ mass scale but still accessible at the LHC, i.e. at the TeV scale. The only light BSM particles are the next-to-lightest SUSY particle, the second-lightest neutralino $\tilde \chi^0_2$, as well a the singlet superfield with its scalar and fermionic components $S$ and $\tilde \chi^0_1$. The latter is the LSP. While $\tilde \chi^0_2$ couples to the SM gauge group (e.g. because it is a bino), the singlet fields only couple very weakly through a small mixture with the other Higgs or neutralino fields, respectively. The production at the LHC therefore proceeds in the coloured sector. Here we assume a gluino pair which then each decays down to the NLSP first, releasing only jets as side-products. The NLSP then decays -- only through the small admixture -- into the NLSP and $S$. While $S$ decays mainly into $b\bar b$, the LSP escapes undetected. 
More precisely, the typical production and decay at the LHC will be
\begin{align}
pp \to \tilde g  \tilde{g} \to qq \bar q \bar q \tilde \chi^0_2 \tilde \chi^0_2 &\to qq \bar q \bar q S S \tilde \chi^0_1 \tilde \chi^0_1\notag \\
&\to 4j+4b+\met\,.
\label{eq:prod_and_decay}
\end{align}
If the LSP is very light, i.e. of $\mathcal O({\rm few ~GeV})$ and the mass gap $\Delta m =  m_{\rm NLSP}-m_{\rm LSP} - m_S$ small, there is only little momentum associated with the escaping LSP. Hence, the signal contains several jets but only very little $\met$. LHC analyses for these kind of scenarios (also including $b$-tags) exist, see e.g. Refs.~\cite{TheATLAScollaboration:2013xia,Khachatryan:2016xim,Aaboud:2018lpl} but can place only relatively loose constraints on the coloured sector compared to typical SUSY searches which require a large amount of $\met$.

The above decay chain contains the leading-order decay of the NLSP. However, since the decay proceeds through a small admixture and is in addition kinematically suppressed by the small available phase space $\Delta m$, it is natural to ask which other decays can be possible, how large they are, and how they affect the detection prospects. These other decay channels constitute of (tree-level) three-body decays as well as (one-loop) radiative decays. While the former will be mediated by an off-shell $Z$ boson, the latter proceeds via loops of charged particles, for instance a charged Higgs and a chargino. These extra decays are therefore given by
\begin{align}
\tilde \chi^0_2 \to \tilde \chi^0_1 Z^*  \to \tilde \chi^0_1 f\bar f \,,
\end{align}
where $Z^*$ denotes an off-shell $Z$ boson, and 
\begin{align}
\tilde \chi^0_2 \to \tilde \chi^0_1 \gamma \,.
\end{align}
In both additional decay modes, the phase space is considerably larger compared to the leading-order two-body decay
due to the larger mass gap between the initial and final state.
In addition, the coupling structure of the new modes is different. As a consequence, they could make for a significant branching ratio. 
Since the NLSP 
is boosted significantly due to the large mass gap, both channels  lead to a potentially detectable $\met$ signal.  \\

In the following, we are going to assess how the additional decay modes of the otherwise stealth scenario can lift the discovery prospects -- and therefore, the bounds on the gluino mass. We will focus on prompt $\tilde \chi^0_2$ decays only. In order to do so, we will assume the following mass hierarchy
\begin{align}
m_{\tilde q} > m_{\tilde g} \gg m_h, M_Z > m_{\tilde \chi^0_2} > m_{S} \gg  m_{\tilde \chi^0_1} \notag \\ 
\text{with}\quad m_{\tilde \chi^0_2} \simeq m_{S} +m_{\tilde \chi^0_1} + {\text (0.5-1)}\,{\rm GeV}\,,
\label{eq:spectrum}
\end{align}
and vary the branching fractions into the two-body, the three-body and the radiative decay freely (from zero to one) in order to access every combination of the three. More precisely, the NLSP, LSP and $S$ masses are set to the values in Tab.~\ref{tab:spectrum}, as inspired by Ref.~\cite{Ellwanger:2014hia}.
\begin{table}[t]
\begin{tabular}{l|c|l|l}
Particle & Mass\,[GeV]\\
\hline
\hline
$\tilde{g}$ & 1100 -- 2000 \\
$\tilde \chi_2^0$ & 89\\
$\tilde \chi_1^0$ & 5\\
$S$ & 83\\
\end{tabular}
\caption{Spectrum of the particles involved in the stealth signal.}
\label{tab:spectrum}
\end{table}
We then test each scenario -- for different gluino masses -- against current LHC analyses.

For the numerical evaluation we 
make use of {\tt Pythia\,8} \cite{Sjostrand:2014zea} in order to generate the Monte Carlo (MC) events with the default parton distribution function {\tt NNPDF\,2.3} \cite{Ball:2012cx}. Hereby we multiply the cross section by a 
$k$-factor as obtained by {\tt NLLfast} \cite{Beenakker:1996ch,Beenakker:1997ut,Kulesza:2008jb,Kulesza:2009kq,Beenakker:2009ha,
Beenakker:2010nq,Beenakker:2011fu}. After that we confront the MC events with the analysis tool \checkmate \cite{Drees:2013wra,Kim:2015wza,Dercks:2016npn} which itself is based on the detector simulator {\tt Delphes\,3} \cite{deFavereau:2013fsa} and the jet reconstruction {\tt Fastjet\,3} \cite{Cacciari:2011ma,Cacciari:2005hq}. \checkmate is a recasting tool which allows the user to test one's model and parameter points against a large number of implemented experimental searches.

We have done scans over the branching ratios BR$(\tilde \chi_2^0 \to \tilde \chi^0_1 S)$, BR$(\tilde \chi_2^0 \to \tilde \chi^0_1 Z^* \to \tilde \chi^0_1 f\bar f )$ and BR$(\tilde \chi_2^0 \to \tilde \chi_1^0 \gamma)$ with the spectrum of Table~\ref{tab:spectrum}. For each point of the grid we have generated 2$\cdot 10^{5}$ MC events.

Once the events for each point are generated we analyse them making use of all the 13 TeV analysis within \checkmate . However, not all the searches are relevant in our purposes and we summarise the ones that are important in Tab.~\ref{tab:lhc_searches}. Now we want to give some details about the searches.

\textbf{Photon(s) + $\mathbf \met$} (1802.03158) \cite{Aaboud:2018doq}: This search is motivated by gauge-mediated supersymmetric breaking (GMSB) models in which final states containing large values of $\met$ and photons are present. Basically their searches can be divided into two regions, the first one is focused on diphoton events with large missing transverse energy while the second one only asks for events with missing energy and the presence of one isolated energetic photon. This search is meant to cover gluino, squark and wino/higgsino production and their subsequent decays to the NLSP that could decay into a gravitino and a photon or a $Z$ boson. In that sense, the signal identical to what we consider here, just in our case instead of the gravitino as the LSP we have the singlino.

\textbf{Multijet + $\mathbf \met$} (1712.02332) \cite{Aaboud:2017vwy}: This search is focused on squark/gluino production and their subsequent decays into quarks giving rise to jets and $\met$. Different signal regions are used which are divided depending on the number of jets they require. In our case, our signal can mimic this kind of searches when the $S$ and the $Z$ decay into quarks so that the relevant signals are jets plus $\met$.

\textbf{Diphoton + $\mathbf \met$} (1606.09150) \cite{ATLASCollaboration:2016wlb}: In this search, events with two photons and large missing energy are required. The motivation is as in Ref.~\cite{Aaboud:2018doq} GMSB where a pair of gluinos is produced decaying to quarks and a neutralino NLSP. This neutralino decays into a gravitino and a photon -- leading to a final state reminiscent of what we are looking for here. However, since this search was performed for low luminosity, $\mathcal{L}=3.2$ fb$^{-1}$, it is less sensitive than Ref.~\cite{Aaboud:2018doq} which searches for the same signal.

\textbf{Leptons + $\mathbf \met$} (1709.05406) \cite{Sirunyan:2017lae}: This search focuses on events with more than two leptons and $\met$ in the final state. It is motivated by the production and subsequent decay into leptons and the LSP of electroweakinos. In our case, higgsino pairs can be produced decaying into the NLSP that could also decay through the $Z$ boson giving the same result. The search is divided into three main regions according to the number of leptons in the final state, two leptons, three leptons or more. 

\begin{table}[tbh]
\begin{tabular}{l|l|l|l}
Reference & Final State & $\mathcal{L}$\,[fb$^{-1}$]\\
\hline
\hline
1802.03158 
\cite{Aaboud:2018doq} & $\geq 1\,\gamma+ {\rm jets} +\met$ & 36.1 \\
1712.02332 \cite{Aaboud:2017vwy} & 2-6 jets + $\met$ &  36.1\\
1606.09150 \cite{ATLASCollaboration:2016wlb} & $2\,\gamma+\met$ & 3.2\\
1709.05406 \cite{Sirunyan:2017lae} & $>2\,\ell+\met$ & 35.5
\end{tabular}
\caption{Summary of the most relevant analyses for our study.
The analyses are referenced by their arXiv number, the third column denotes the final
state topology, and the fourth column shows the total integrated luminosity. All analyses have been performed with 13\,TeV of centre-of-mass energy.}
\label{tab:lhc_searches}
\end{table}

\begin{figure*}
\centering
\includegraphics[width=.49\linewidth]{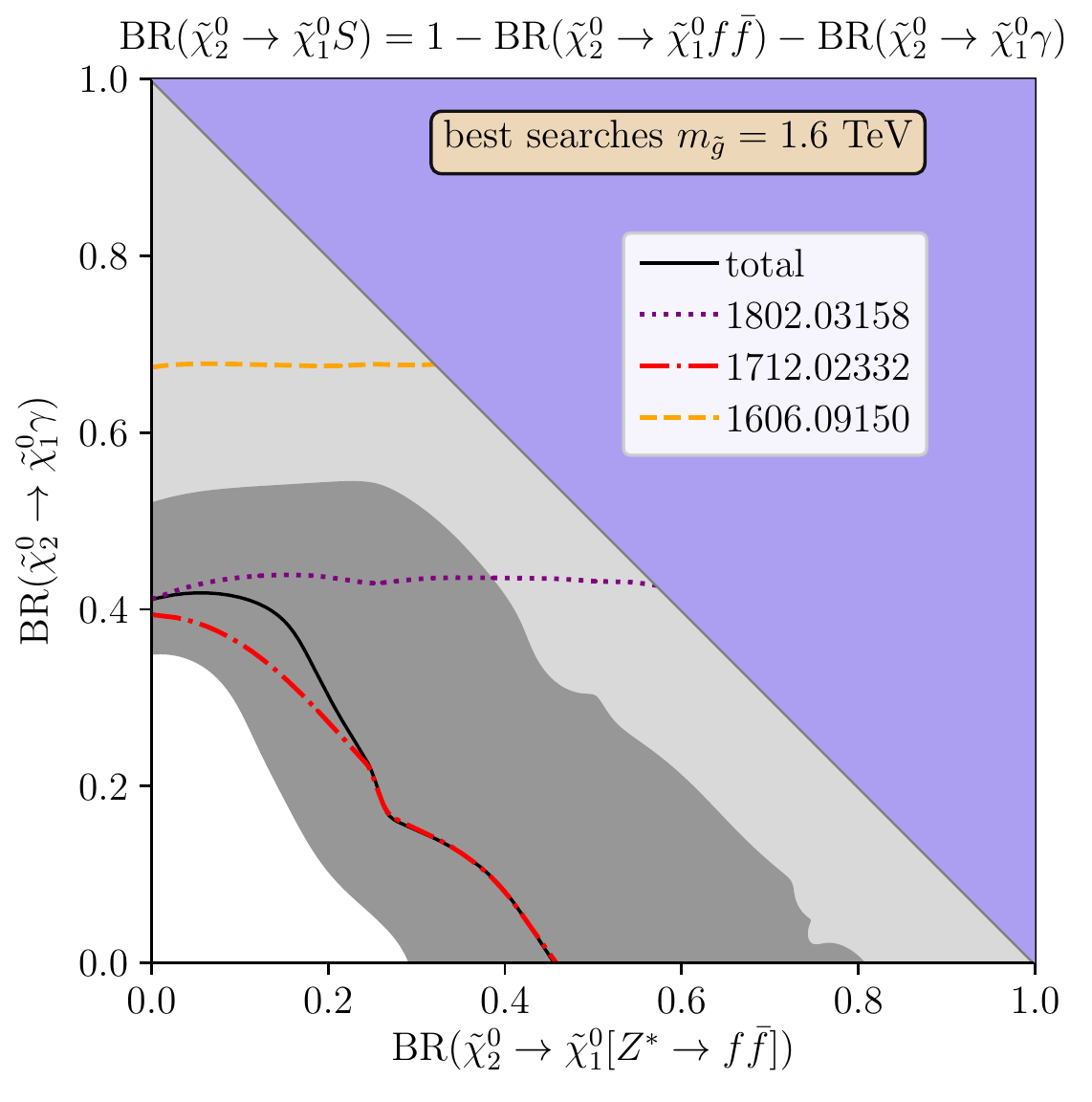}
\includegraphics[width=.49\linewidth]{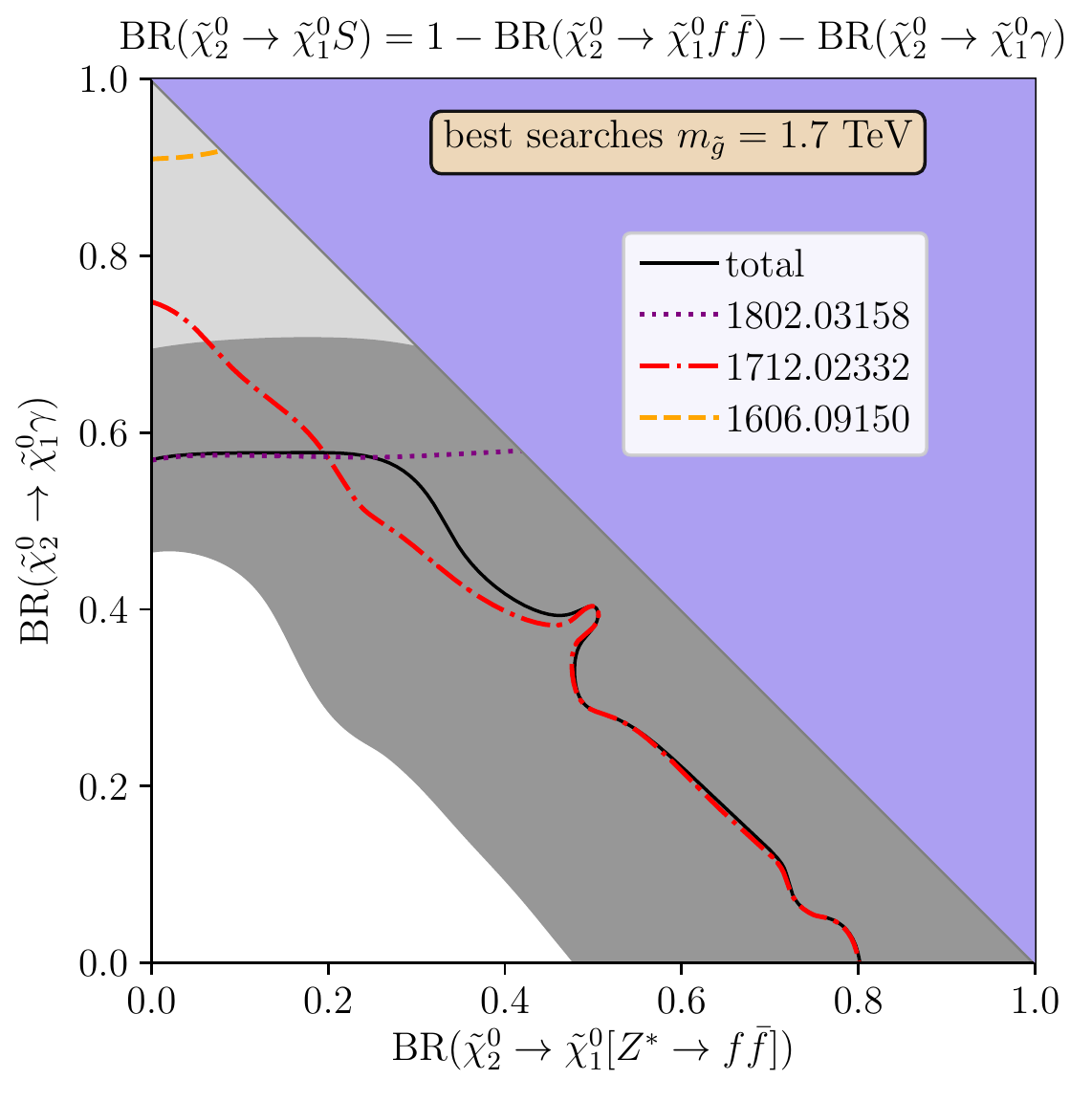}
\caption{Excluded model space depending on the branching ratio of $\tilde \chi^0_2$ into the the three-body and the radiative two-body final state. In the left-hand plot, we fix $m_{\tilde g} = 1.6\,$GeV while we set $m_{\tilde g} = 1.7\,$GeV in the right-hand pane. We display the respective most restraining analyses in dots and dashes. They are summarised in Tab.~\ref{tab:lhc_searches}. The total exclusion is shown in solid black. The dark grey shading corresponds to regions with $1.5 < r < 0.67$ and therefore ambiguous exclusion, whereas the light grey regions are ruled out.}
\label{fig:simpmodels}
\end{figure*}

\begin{figure}
\centering
\includegraphics[width=\linewidth]{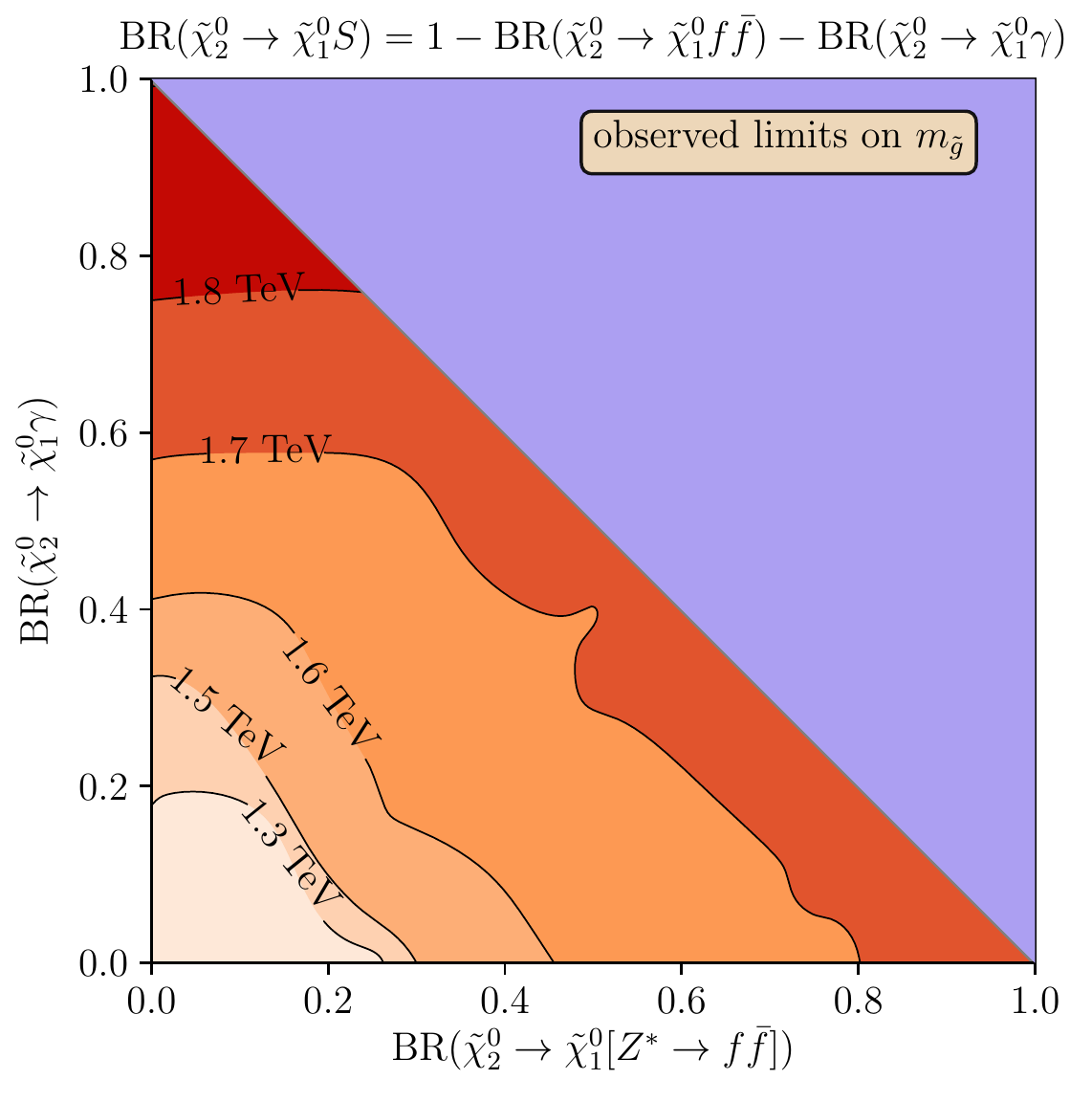}
\caption{Upper bounds on the gluino mass depending on the branching ratio of $\tilde \chi^0_2$ into the the three-body and the radiative two-body final state. }
\label{fig:simpmodelsgeneral}
\end{figure}

In order to determine whether a point is excluded by a search or not, we compare the estimate of signal events with the observed limit at 95\% C. L. of the search in the following way,
\begin{eqnarray}
r=\frac{s-1.96\cdot \Delta s}{s^{95}_{\rm exp}}.
\end{eqnarray}
$s$ denotes the number of signal events, $\Delta s$ the uncertainty of MC events that we consider to be only the statistical uncertainty, $\Delta s = \sqrt{s}$. This quantity is calculated for every signal region of every search. Then, in order to calculate the best exclusion limit we choose the `best' signal region which we define as the one with the best expected exclusion potential. As a result, the total exclusion limit could be weaker than the limits from a single signal region. In \checkmate it is not possible to combine searches, so the limits which we calculate are conservative. One can define a point as excluded when the $r$-value is greater than $r > 1$. However, as we do not control higher-order corrections or systematic errors we define a region where exclusion is inconclusive. This region is the one between $0.67<r<1.5$. When one of the points is placed in this region we cannot tell if it is excluded or not since a fluctuation in the estimate of the signal number of events due to missing correction could change the result. According to this we define a point as allowed when it presents a value $r<0.67$ and excluded when $r>1.5$. 

\subsection{Gluino searches}
In Fig.~\ref{fig:simpmodels}, the corresponding exclusion limits for gluino masses of $m_{\tilde g}=1.6$ TeV (left) and $m_{\tilde g}=1.7$ TeV (right) are depicted. 
In both panels the exclusion contour line is plotted as a function of the different branching ratios of the neutralino NLSP, $\tilde \chi_2^0$. In the $x$-axis we plot the branching ratio into $Z$ boson, BR$(\tilde \chi^0_2 \to \tilde \chi^0_1 Z^*  \to \tilde \chi^0_1 f\bar f )$, while in the $y$-axis we show the branching ratio into a photon and the neutralino LSP, BR$(\tilde \chi^0_2 \to \tilde \chi^0_1 \gamma )$. The third branching ratio, corresponding to $\tilde \chi^0_2 \to S \tilde \chi^0_1$, is given for each point as  
\begin{eqnarray}
{\rm BR}(\tilde \chi^0_2 \to \tilde \chi^0_1 S )=1-{\rm BR}(\tilde \chi^0_2 \to \tilde \chi^0_1 f\bar f ) \notag \\
-{\rm BR}(\tilde \chi^0_2 \to \tilde \chi^0_1 \gamma).
\label{eq:branchingratios}
\end{eqnarray}
We have covered in purple colour the non-physical area where the total sum of branching ratios is greater than 100\%.
 
We see in Fig.~\ref{fig:simpmodels} that the exclusion lines from the analyses Refs.~\cite{ATLASCollaboration:2016wlb,Aaboud:2018doq} are horizontal, meaning that they only depend on the branching fraction into the photonic final state as expected. Ref.~\cite{Aaboud:2017vwy}, in turn, tags $\met$ and jets,  which is provided by both final states on the $x$- and $y$-axis, leading to almost diagonal lines. 
Correspondingly, for low BR$(\tilde \chi^0_2 \to \tilde \chi^0_1 \gamma )$, the jets+$\met$ search sets the best exclusion limits, while for large branching ratio into photons, the photonic searches are most efficient, which can be seen in the right-hand plot of Fig.~\ref{fig:simpmodels}.
This is also seen in Fig.~\ref{fig:simpmodelsgeneral} where we compile the bounds on the gluino masses as a function of the three branching ratios. It is seen that,  for BR$(\tilde \chi^0_2 \to \tilde \chi^0_1 S) \simeq 100\%$, only gluino masses up to 1.2\,TeV can be excluded, this quickly changes with increasing branching ratio of the alternative decays, leading to bounds up to 1.8\,TeV for the case of BR$(\tilde \chi^0_2 \to \tilde \chi^0_1 \gamma) \simeq 100\%$.

\subsection{Higgsino searches}
Apart from the coloured sector, also the detection prospects for electroweakinos can be reduced significantly by a compressed NLSP decay. Although the corresponding searches look for multilepton final states, the signal regions are complemented with rather tight $\met$ cuts in order to enhance the separation from the background. Consider, for instance, the CMS analysis of Ref.~\cite{Sirunyan:2017lae}: out of many signal regions (depending on the number and signs of leptons), only a few tag missing transverse momentum as low as 50\,GeV -- most are a lot tighter. 

In a natural SUSY environment, featuring rather light higgsinos, the higgsinos and their decay products could therefore be hidden if they decay down to a (gauge) boson and the NLSP, with its subsequent stealth decay. In the following we briefly show how this higgsino-stealth scenario is washed out by the effect of the alternative decay modes considered before.

For that purpose we have performed a scan over the mass of the higgsinos. In this scenario we consider the direct production of the higgsinos and their subsequent decays into the second lightest neutralino, $\tilde \chi_2^0$,
\begin{eqnarray}
pp \to \tilde \chi^\pm_1  \tilde \chi^0_{3,4} \to W^\pm \, \chi^0_{2} \, Z/h \, \chi^0_{2} \notag \\
pp \to \tilde \chi^\pm_1  \tilde \chi^\mp_1 \to W^\pm \, \chi^0_{2} \,W^\pm \, \chi^0_{2}  
\label{eq:prodhiggsino}
 \\
pp \to \tilde \chi^0_{3,4}  \tilde \chi^0_{3,4} \to Z/h\,  \chi^0_{2}\, Z/h \, \chi^0_{2}\notag 
\end{eqnarray}
that will decay as we described above. We take the leading order cross section from {\tt Pythia8} and apply a conservative flat $\mathcal{K}$-factor of 20\%. Here we perform the scan over the branching ratios as in the case of the gluino, and we also scan over the lightest neutral higgsino mass, $m_{\tilde \chi_3^0}$, while we consider the following hierarchy $m_{\tilde \chi_4^0}=m_{\tilde \chi_1^\pm}=m_{\tilde \chi_3^0} + 5$ GeV. For the numerical evaluation we proceed as in the gluino case. We have split the scans in two different scenarios depending on the decay of the neutralino, $\chi^0_3$ into a $Z$ boson ($\chi_3^0\to \chi_2^0 Z$) or a Higgs boson ($\chi_3^0\to \chi_2^0 h$).

\begin{figure*}
\centering
\includegraphics[width=.49\linewidth]{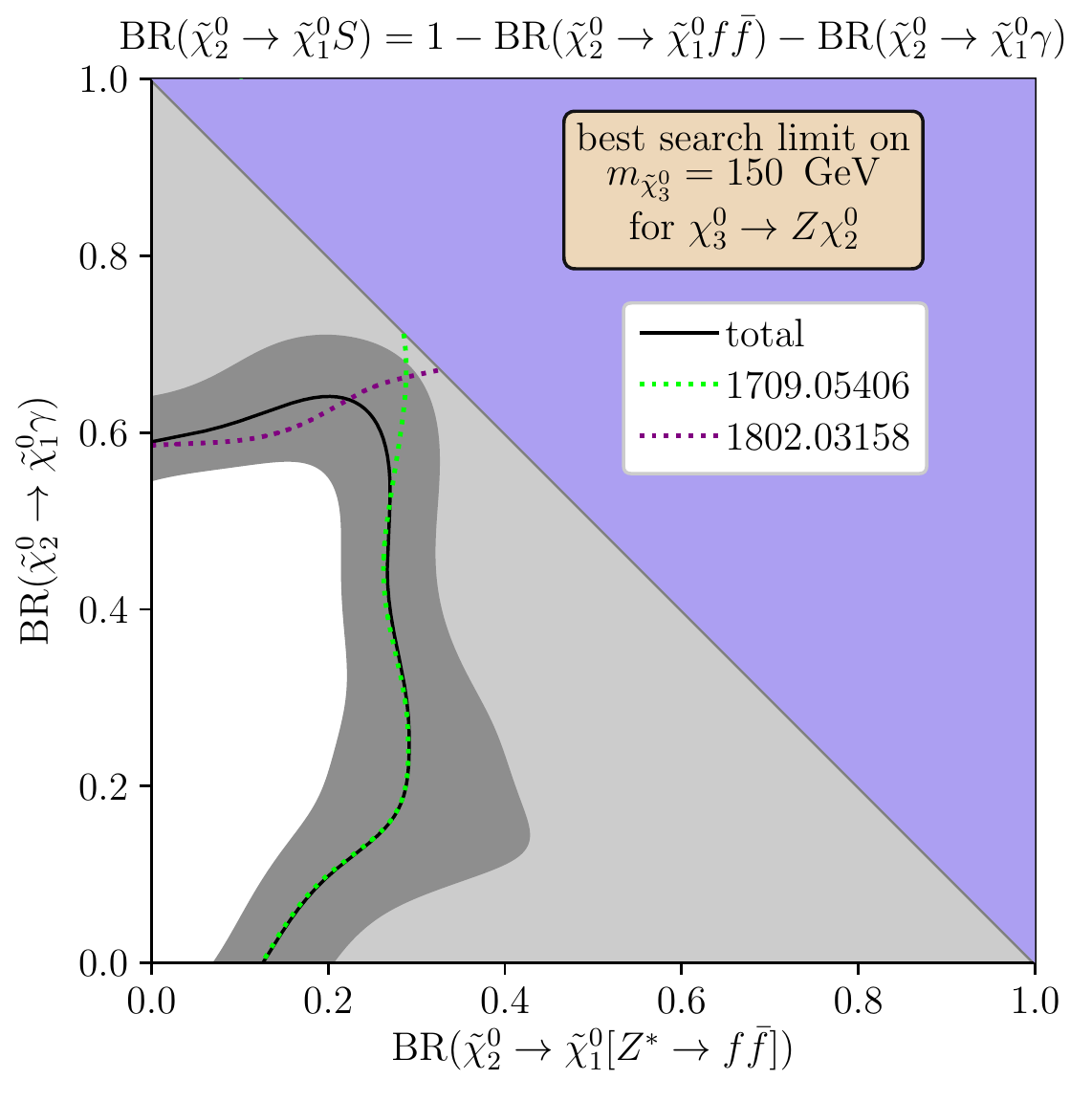}
\includegraphics[width=.49\linewidth]{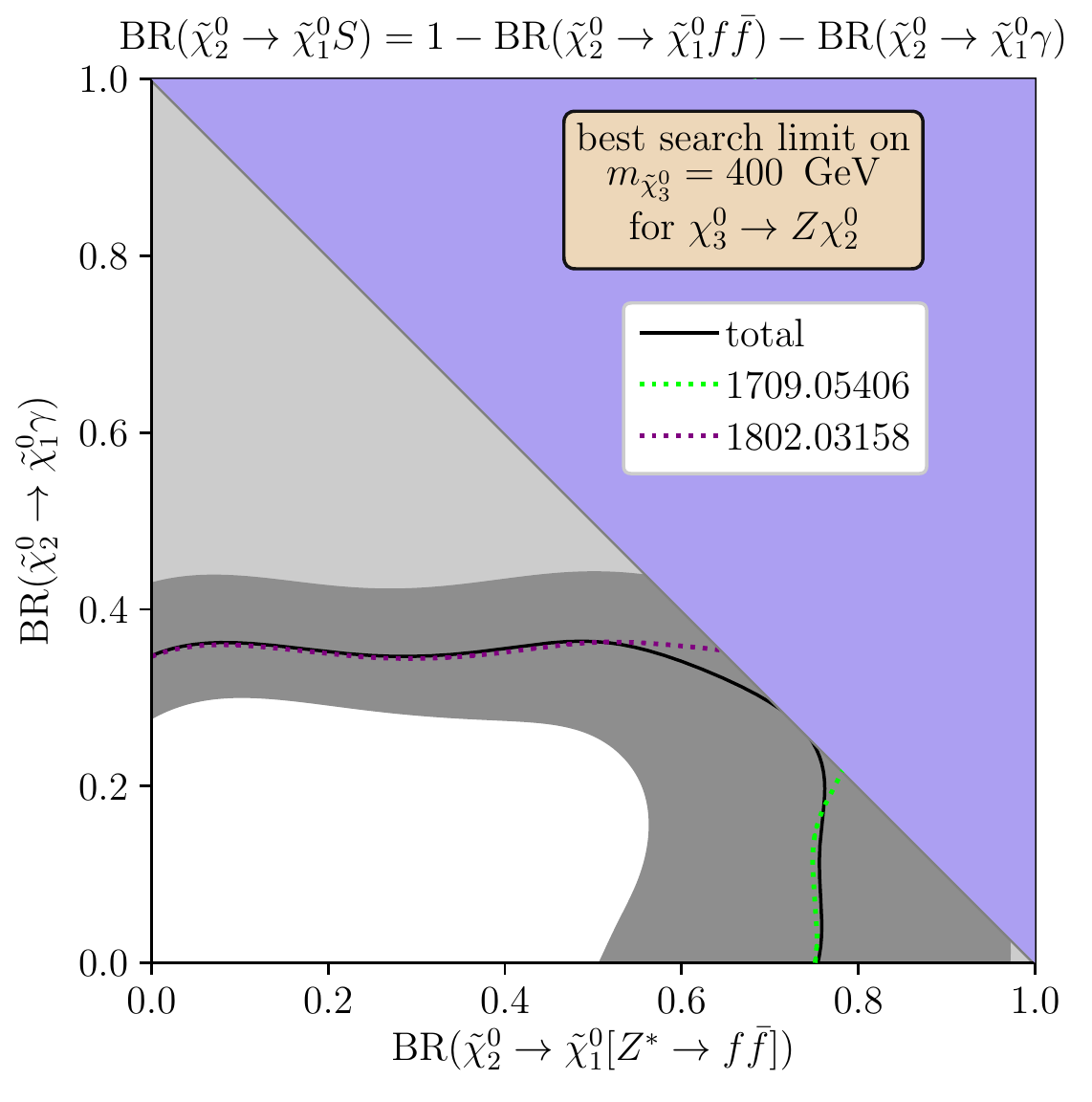}
\caption{Excluded model space depending on the branching ratio of $\tilde \chi^0_2$ into the the three-body and the radiative two-body final state assuming BR($\chi_3^0\to Z \chi_2^0$)=1. In the left-hand plot, we fix $m_{\tilde \chi_3^0} = 150\,$GeV while we set $m_{\tilde \chi_3^0} = 400\,$GeV in the right-hand pane. We display the respective most restraining analyses in dashes. They are in this case the multilepton plus $\met$ analysis~\cite{Sirunyan:2017lae} (green) and the photons plus $\met$ search~\cite{Aaboud:2018doq}. The total exclusion is shown in solid black. The dark grey shading corresponds to regions with $1.5 < r < 0.67$ and therefore ambiguous exclusion, whereas the light grey regions are ruled out. }
\label{fig:higgsinosZ}
\end{figure*}

In Fig.~\ref{fig:higgsinosZ} we can see the exclusion limits for the higgsino-like neutralino masses of $m_{\tilde \chi_3^0}=150$ GeV (left) and $m_{\tilde \chi_3^0}=400$ GeV (right) assuming that the third neutralino decays totally into the second neutralino, $\chi_2^0$, and the $Z$ boson. The axes correspond as in the gluino case to the branching ratio into a $Z$ boson and into a photon and the neutralino LSP, while the corresponding branching ratio into the singlet scalar and neutralino LSP is obtained with Eq.~\eqref{eq:branchingratios}. As we did with the gluino plots, we have covered in purple colour the non-physical area for the branching ratios.

In the left panel of Fig.~\ref{fig:higgsinosZ} the exclusion limits for a higgsino mass of $m_{\chi_3^0}$ = 150 GeV is depicted. The total exclusion rate is depicted as a solid black line, and it is constructed from the different searches. There are two sensitive searches in  this scenario that are the multileptonic analysis of Ref.~\cite{Sirunyan:2017lae} and the photonic search of Ref.~\cite{Aaboud:2018doq}. The first one, depicted as a green dashed line, is able to exclude all the points which branching ratio into $Z$ boson greater than 20-30\%. It is almost insensitive to the other branching ratios except for large values of the decay into a singlet, $S$, while the photonic decay is low. In this case the exclusion can cover smaller values of the branching ratio into a $Z$ up to 15\%. This search is really powerful in the low mass region since the requirements of the search are designed to prove these electroweakino masses and also due to the large cross section. The second search, shown as a dashed purple line, is only able to test the regime of large values of photonic decays, larger than 60\%. One has to say that in this scenario the photonic search is less sensitive since the cuts applied in the analysis required large values of transverse variables that are typical from particles with larger masses. This search is insensitive to the other branching ratios and the value from which it is sensitive is almost constant, as it happens for the gluino case. The total exclusion area for a neutralino of mass $m_{\chi_3^0}=150$ GeV is rounded by a solid black line. The exclusion power is really high since the allowed region left after applying the analysis is reduced to large values of the decay into singlets.

In the right side of Fig.~\ref{fig:higgsinosZ} we show the exclusion limit for a higgsino mass of $m_{\chi_3^0}$ = 400 GeV. The colour code is the same as in the other case. The leptonic search here is less powerful since it can only constrain large branching ratios into $Z$ bosons, ie. BR($\tilde\chi_2^0 \to \tilde{\chi}_1^0[Z^*\to f\bar{f}]$) $>$ 75\%. The photonic search for this mass seems to be most powerful constraining branching ratios greater than 35\%. This fact is due to the strong cuts imposed in the photonic analysis \cite{Aaboud:2018doq} that require large values of transverse variables typical from larger masses.

\begin{figure*}
\centering
\includegraphics[width=.49\linewidth]{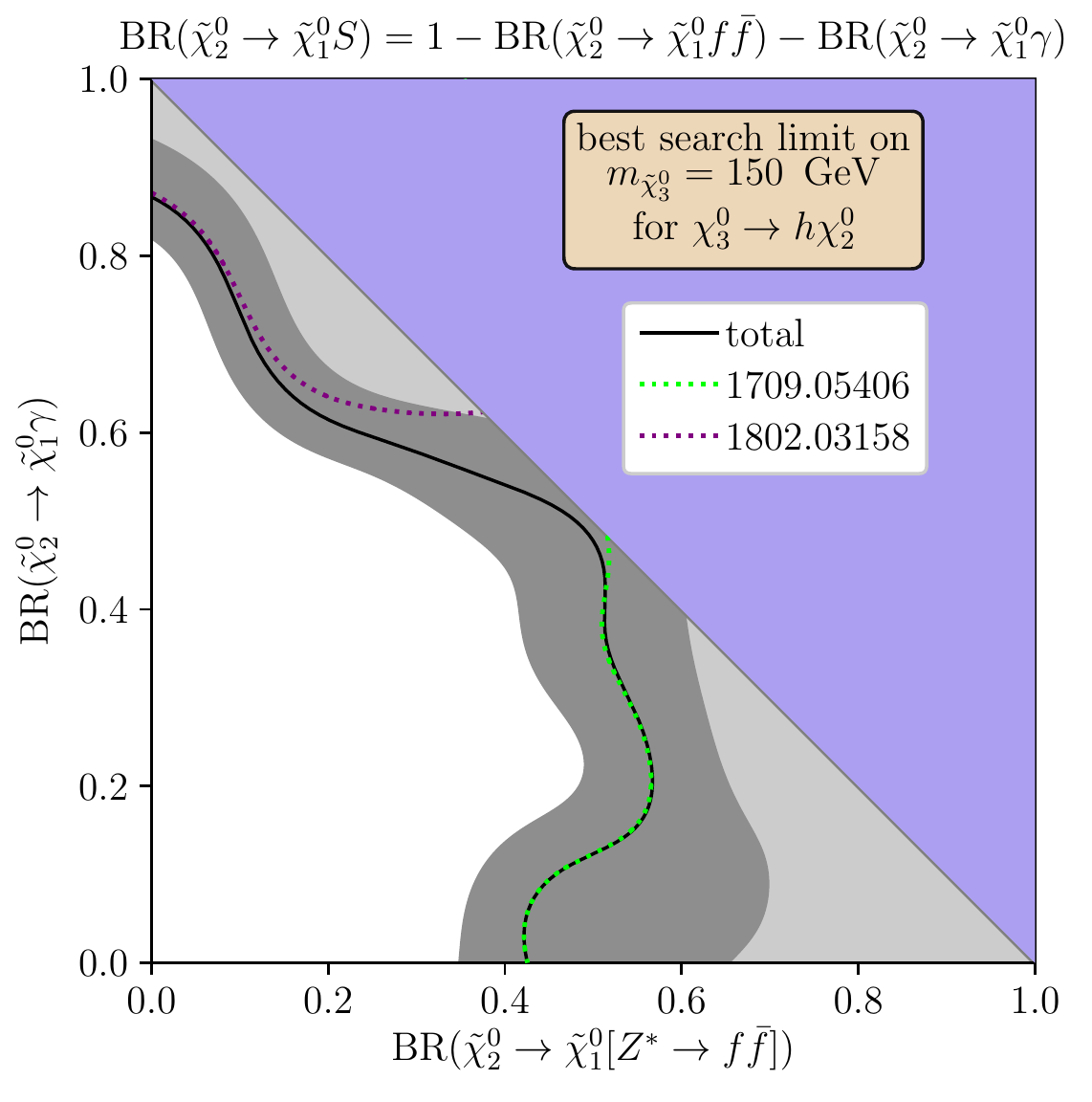}
\includegraphics[width=.49\linewidth]{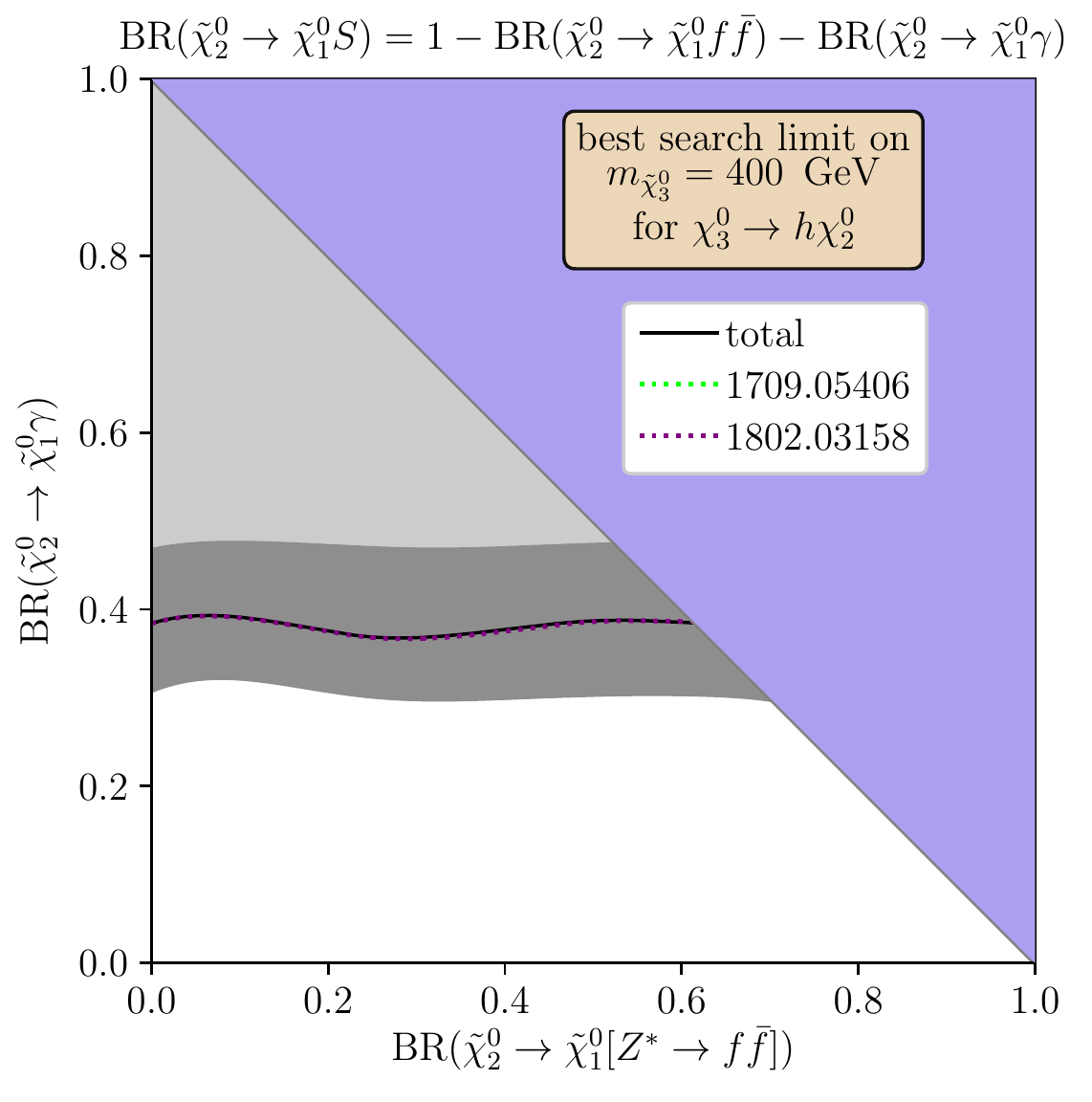}
\caption{Excluded model space depending on the branching ratio of $\tilde \chi^0_2$ into the the three-body and the radiative two-body final state assuming BR($\chi_3^0\to h \chi_2^0$)=1. In the left-hand plot, we fix $m_{\tilde \chi_3^0} = 150\,$GeV while we set $m_{\tilde \chi_3^0} = 400\,$GeV in the right-hand pane. We display the respective most restraining analyses in dashes. They are in this case the multilepton plus $\met$ analysis~\cite{Sirunyan:2017lae} (green) and the photons plus $\met$ search~\cite{Aaboud:2018doq} (purple). The total exclusion is shown in solid black. The dark grey shading corresponds to regions with $1.5 < r < 0.67$ and therefore ambiguous exclusion, whereas the light grey regions are ruled out. }
\label{fig:higgsinosH}
\end{figure*}

In Fig.~\ref{fig:higgsinosH} the same scenario is shown as in Fig.~\ref{fig:higgsinosZ} but assuming that the third neutralino decays totally into Higgs bosons, BR($\tilde{\chi}_3^0\to \tilde{\chi}_2^0 h$) = 100\%. In the left panel of Fig.~\ref{fig:higgsinosH} the results for a mass $m_{\tilde{\chi}_3^0}$ = 150 GeV are depicted. As we can see the exclusion limit is weaker than in the previous case of Fig.~\ref{fig:higgsinosZ} where we assume decay into $Z$ bosons for the third neutralino. This is the reason why here the leptonic search is not as powerful as in the previous case. Since the third neutralino decays into a Higgs boson the leptonic rate is smaller, now only the second neutralino provides leptonic events. In this scenario the photonic search is also weaker than in the previous one for larger values of the decay into singlets. In the right panel of Fig.~\ref{fig:higgsinosH} we present the results for the same scenario for a mass of the third neutralino of $m_{\tilde{\chi}_3^0}$ = 400 GeV. In this case the only search that is able to constrain this scenario is the photonic one. As in the case where the third neutralino decays into $Z$ bosons the limit is constant and fixed in a value of BR($\tilde{\chi}_3^0\to\tilde{\chi}_2^0$) $>$ 40\%. However, now we do not have the exclusion area given by the leptonic searches. This fact is because now the third neutralino decays into a Higgs boson giving fewer leptonic events.

\begin{figure*}[ht!]
\centering
\includegraphics[width=.49\linewidth]{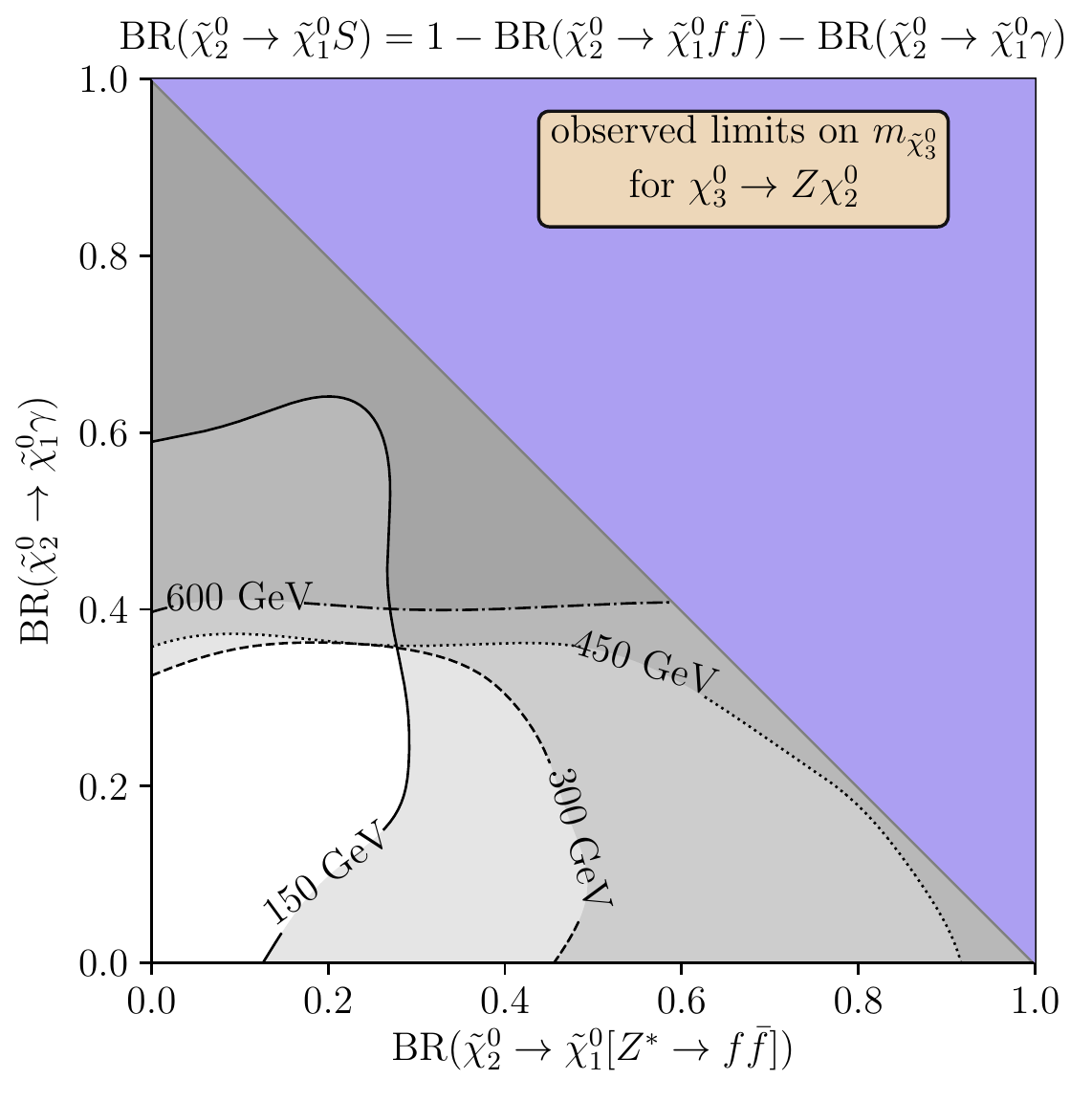}
\includegraphics[width=.49\linewidth]{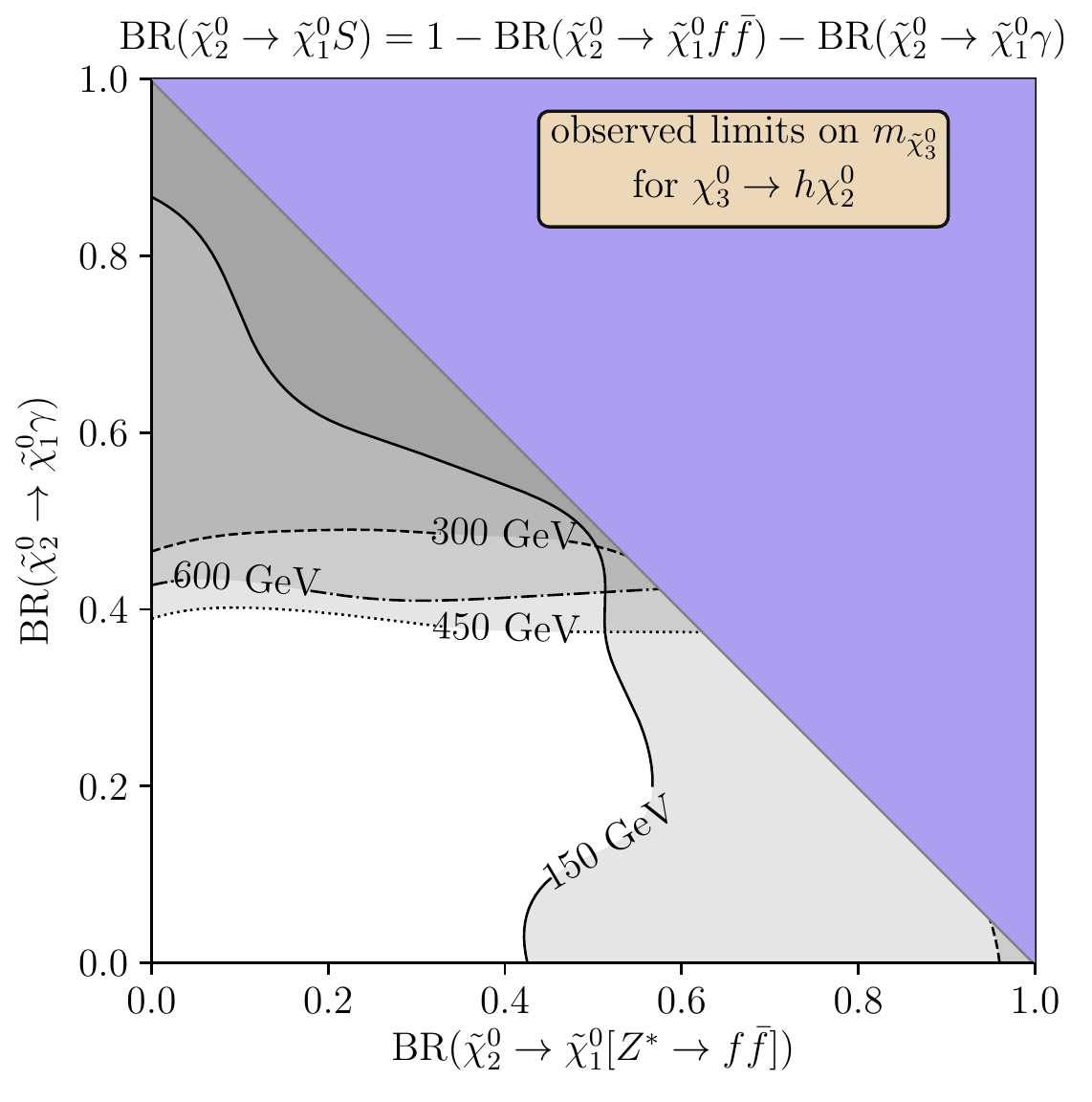}
\caption{Upper bounds on the higgsino mass, $m_{\tilde \chi_3^0}$, depending on the branching ratio of $\tilde \chi^0_2$ into the the three-body and the radiative two-body final state for BR$(\tilde{\chi}_3^0\to Z \tilde{\chi}_2^0)$ = 100\% (left) and BR$(\tilde{\chi}_3^0\to h \tilde{\chi}_2^0)$ = 100\% (right). }
\label{fig:simpmodelsgeneralhiggsino}
\end{figure*}

The limits for different higgsino masses in both scenarios are summarised in Fig.~\ref{fig:simpmodelsgeneralhiggsino}. In the left panel of Fig.~\ref{fig:simpmodelsgeneralhiggsino}, the scenario of BR($\tilde\chi_3^0\to Z \tilde\chi_2^0$) is depicted. We can see  that with increasing neutralino mass the leptonic search loses sensitivity until we reach masses greater than $m_{\tilde \chi_3^0}$ $>$ 450~GeV where this search becomes totally insensitive. On the contrary the photonic search becomes more stringent once we reach masses greater than $m_{\tilde \chi_3^0}$ $>$ 200~GeV. These searches can constrain branching ratios of about BR($\tilde{\chi}_2^0\to\tilde{\chi}_1^0 \gamma$) $\gtrsim$ 35-40\% in the range of masses $m_{\tilde\chi_3^0}$ = 300-600~GeV. In the case of BR($\tilde\chi_3^0\to h \tilde\chi_2^0$), shown in the right panel of Fig.~\ref{fig:simpmodelsgeneralhiggsino}, the leptonic search is much less efficient since it drops quickly for masses greater than $m_{\tilde\chi_3^0}$ $>$ 150~GeV. This is due to the lack of leptonic events since the third neutralino decays into the Higgs boson only. For larger masses the photonic search becomes the most stringent one and it is as sensitive as in the previous case, excluding BR($\tilde{\chi}_2^0\to\tilde{\chi}_1^0 \gamma$) $\gtrsim$ 40-45\% for masses $m_{\tilde\chi_3^0}$ = 300-600 GeV.


It is important to note here that in these stealth spectra where the only production comes from the higgsino sector, a usual configuration where the branching ratio of the second lightest neutralino into singlets that do not exceed values greater than 70-80\% are totally invisible to the LHC. So a typical second lightest neutralino that decays mainly into a singlet and the LSP could be totally invisible even if the higgsino masses are close to the LEP limit.



\section{Stealth scenario in SUSY models}
\label{sec:examples}

We  now turn to the discussion of concrete (SUSY) models which in principle provide all ingredients for a stealth scenario.

\subsection{St\"uckelberg extension of the MSSM}
\label{sec:stuckel}
 We start with a model which was to our knowledge not yet been discussed in this context. As we will see, there are good reasons for this because the three-body decays are crucial and rule out this idea immediately. Nevertheless, it might deal as a nice example to show how dangerous it is to rely on the calculation of only two-body decays. The model which we want to discuss briefly is the minimal St\"uckelberg extension of the MSSM \cite{Kors:2004ri} which extends the SM gauge sector by a new Abelian gauge group $U(1)_X$.  The superpotential is just the one of the standard MSSM
\begin{align}
W_{\rm MSSM} &= Y_u \hat H_u \hat Q \hat u + Y_d \hat H_d \hat Q \hat d + Y_e \hat H_d \hat L \hat e +  \mu \hat H_u \hat H_d  \notag \\
& = W_Y +  \mu \hat H_u \hat H_d
\label{eq:w_stueckel}
\end{align}
where all fields are uncharged under $U(1)_X$. 
The additional particles compared to the MSSM are a vector superfield $\hat{B}'$ and a gauge singlet $\hat \rho$.  Even if $\hat \rho$ is a complete singlet, it can nevertheless generate a mass term for the new gauge boson $B'$. The St\"uckelberg Lagrangian is given by
\begin{equation}
\mathcal{L}_{St} = \int d \Theta^2 d \bar{\Theta}^2 (m_1 \hat B' + m_2 \hat B + \hat \rho + \hat{ \bar{ \rho}})^2
\end{equation}
where $\hat B$ is the vector superfield of the hypercharge group. The new physical states are two additional neutralinos from the gauge eigenstates $\tilde{B}'$, $\tilde{S}$, one CP-even scalar which mixes with the CP-even Higgs from the MSSM but which mainly consists of $\mathcal{R}(\rho) \equiv \phi_\rho$, one new gauge boson $Z'$ which is mainly a $B'$. Up to small mixings, the masses of the bosonic states are given by 
\begin{equation}
m^2_{\phi_\rho} \simeq m_1^2 + m_2^2 , \hspace{1cm} m^2_{Z'} \simeq m_1^2\,,
\end{equation}
while the $Z-Z'$ mixing is proportional to
\begin{equation}
\epsilon = \frac{m_2}{m_1}\,.
\end{equation}
Also the mixing between $\phi_\rho$ and the other CP even scalar is  $O(\epsilon)$. \\ 
The neutralino mass matrix for this model reads in the basis $(\tilde{\rho},\tilde{B}', \tilde{B}, \tilde{W}^0, \tilde{H}_d^0, \tilde{H}_u^0)$
\begin{equation}
M_{\tilde{\chi}^0} = \left(
\begin{array}{cccccc}
 0 & m_1 & m_2 & 0 & 0 & 0 \\
 m_1 & M_4 & 0 & 0 & 0 & 0 \\
 m_2 & 0 & M_1 & 0 & -\frac{g_1 v_d}{2} \
& \frac{g_1 v_u}{2} \\
 0 & 0 & 0 & M_2 & \frac{g_2 v_d}{2} & -\frac{g_2 v_u}{2} \\
 0 & 0 & -\frac{g_1 v_d}{2} & \frac{g_2 v_d}{2} & 0 & -\mu  \\
 0 & 0 & \frac{g_1 v_u}{2} & -\frac{g_2 v_u}{2} & -\mu  & 0 \\
\end{array}
\right)
\end{equation}
Here, $M_i$ are the gaugino soft SUSY-breaking terms.
Considering only the $3\times 3$ submatrix of $\tilde{S}$,$\tilde{B}'$, $\tilde{B}$ in the limit $m_2 \to 0$, one finds that the three eigenvalues are
\begin{equation}
 M_1\,, \hspace{1cm} \frac12\left(M_4 \pm \sqrt{M^2_4 + 4 m_1^2} \right)\,.
\end{equation}
Thus, for $M_4 \gg m_1$ one state becomes very light. So, we see that without much tuning one can find a kinematic configuration with 
\begin{eqnarray}
m_{\rm NLSP} = m_{\tilde{\chi}^0_2} \simeq M_1 \lesssim m_Z \,,\\
m_{\rm LSP} = m_{\tilde{\chi}^0_1} \simeq 0 \,,\\
m_{h_1} \simeq m_{\phi_\rho} \simeq  m_{\tilde{\chi}^0_2}- m_{\tilde{\chi}^0_1}\,.
\end{eqnarray}
However, the vertex responsible for the NLSP two-body decay is highly suppressed  because the $\rho$ field interacts neither via gauge interactions nor superpotential terms. Therefore, the vertex is proportional 
to the mixing of the involved scalar with the Higgs doublets. An additional suppression comes with the small Higgsino fraction of the mainly $\tilde{\rho}$-like LSP. The interaction strength can be approximated as
\begin{equation}
V_{\tilde{\chi}^0_2 \tilde{\chi}^0_1 h_1} \sim \frac{\epsilon^2 g_1^3 \tan\beta v^2}{4 M_1 \mu (1 + \tan\beta)}\,.
\end{equation}
One can compare this now with the ${\tilde{\chi}^0_2 - \tilde{\chi}^0_1 - Z}$ vertex which triggers the three-body decays of the NLSP via an off-shell $Z$-boson. This vertex is also suppressed by the Higgsino fraction of the LSP and the second suppression factor is due to the Bino-Higgsino mixing. However, this suppression is not propotional to $\epsilon$ but can be much weaker. All in all, we find
\begin{equation}
V_{\tilde{\chi}^0_2 \tilde{\chi}^0_1 Z} \sim -\frac{g_1^2 m_1 \epsilon \tan^2\beta v^2 (g_2 \cos\theta_W + g_1 \sin\theta_W)}{8 M_1 \mu^2 (1 + \tan^2\beta ) }\,.
\end{equation}
Thus, the ratio of both is 
\begin{equation}
\frac{V_{\tilde{\chi}^0_2 \tilde{\chi}^0_1 h_1} }{V_{\tilde{\chi}^0_2 \tilde{\chi}^0_1 Z}} \sim \frac{m_1 \tan\beta}{\epsilon \mu}\,.
\end{equation}
Since the usual suppression of three-body decays compared to two-body decays is also compenstated by the much larger phase space for the three-body decays in this case, one can expect that the three-body partial width clearly dominates. Since $\epsilon$ can be at most $O(0.01)$ because of precision data and current $Z'$ searches for such light $Z'$-bosons \cite{Feldman:2006wb},  one would need Higgsino masses in the multi-TeV range to make the two-body decays at least competitive with the three-body decays. For Higgsino masses of a few hundred GeV, the branching ratio of the two-body decay is only of the level of $10^{-4}$--$10^{-5}$. Therefore, we consider this scenario as not very attractive and turn directly to a more interesting example.

\subsection{NMSSM}
\label{sec:NMSSM}

Let us consider the MSSM extended by a singlet superfield $\hat S$, commonly known as the NMSSM.
We are going to investigate a slightly altered version, where the main difference w.r.t. more common versions of the NMSSM is that we explicity allow for a $\mathbb Z_3$-breaking $\mu$-term.
The superpotential then reads
\begin{align}
W &= \lambda \hat H_u \hat H_d \hat S + \frac{\kappa}{3}\kappa \hat S^3 +  \mu \hat H_u \hat H_d 
+  W_Y \,,
\end{align}
where $W_Y$ contains the standard Yukawa interactions as in the MSSM, cf. Eq.~(\ref{eq:w_stueckel}). In addition to the MSSM  soft SUSY-breaking terms, we consider the following terms:
\begin{align}
- \mathcal L_{\rm soft} &\supset \left(T_\lambda H_u H_d S + \frac{T_\kappa}{3}  S^3  + \frac{B_s}{2} S^2 + \xi_s S + \text{h.c.} \right) \notag \\ &+ m_s^2 |S|^2\,,
\end{align}
where we defined the trilinear soft terms
$ T_\lambda = A_\lambda \lambda \,,\, T_\kappa = A_\kappa \kappa$.
After electroweak symmetry breaking, the scalar doublets $H_{u,d}$ as well as the singlet scalar $S$ receive vacuum expectation values (VEVs) $v_{u,d,S}$ according to $\langle \phi_i\rangle = v_i/\sqrt{2}$. Therefore, the `effective $\mu$-term' reads $\mu_{\rm eff} = \mu + \frac{ \lambda v_S}{\sqrt{2}}$. The ratio of the doublet VEVs is defined as $\tan\beta = v_u/v_d$.

In addition to the MSSM spectrum, the extra singlet superfield leads to one more CP-even and one CP-odd scalar as well as one additional neutralino. Despite a small admixture by the doublet states, we denote the additional CP-even (odd) scalars as $S$ ($A_S$).

The region in parameter space which we are about to consider is very much inspired by Ref.~\cite{Ellwanger:2014hia}, with the difference that we use the additional freedom which we obtained by adding the $\mu$-term to (i) lift the mass of $A_S$ w.r.t. $m_{S}$\footnote{The singlet mass $m_S$ is \emph{not} the same as the soft mass term $m_s$. } and (ii) add a $\lambda$-independent term to $\mu_{\rm eff}$, and therefore the higgsino mass. In order to arrive in a stealth parameter region, we use the hierarchy 
of Eq.~(\ref{eq:spectrum}) 
where $\tilde \chi^0_2$ is bino-like and $\tilde \chi^0_1$ singlino-like. 
This corresponds to the situation in Fig.~\ref{fig:massconfig} where $\tilde \chi^0_{1\,(2)}$ is the (N)LSP.
Correspondingly, the leading-order production and decay chain will be the same as in Eq.~(\ref{eq:prod_and_decay}).
So far, this is exactly the situation described in Ref.~\cite{Ellwanger:2014hia}.\footnote{Note that the constellation discussed in Refs.~\cite{Ellwanger:2014hca,Titterton:2018pba}, although very similar in principle, differs in an important detail: in that case, the NLSP decay ends in the LSP and a SM Higgs instead of $S$. Correspondingly, searches for two SM Higgs bosons and hard jets become sensitive.
}

%

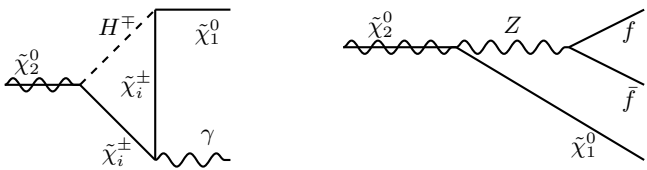
\begin{figure}
\begin{tikzpicture}
\begin{scope}[thick] 

\draw[-] (0,0)--(1,0);
\draw[photon] (0,0)--(1,0);
\draw[-,dashed] (1,0) -- (2,1);
\draw[-] (1,0) -- (2,-1);
\draw[-] (2,-1) -- (2,1);
\draw[-] (2,1) -- (3,1);
\draw[photon] (2,-1) -- (3,-1);
\node[black] at (0.3,0.3) {{$\tilde \chi^0_2$}};
\node[black] at (2.7,0.7) {{$\tilde \chi^0_1$}};
\node[black] at (2.7,-0.7) {{$\gamma$}};
\node[black] at (1.5,0.8) {{$H^\mp$}};
\node[black] at (1.5,-0.9) {{$\tilde \chi_i^\pm$}};
\node[black] at (1.75,0) {{$\tilde \chi_i^\pm$}};

\draw[-] (4.5,0.5) -- (6,0.5);
\draw[photon] (4.5,0.5) -- (6,0.5);
\draw[photon] (6,0.5) -- (7.5,0.5);
\draw[-] (7.5,0.5) -- (8.5,1);
\draw[-] (7.5,0.5) -- (8.5,0);
\draw[-] (6,0.5) -- (8.5,-1);
\node[black] at (5,0.8) {{$\tilde \chi^0_2$}};
\node[black] at (7.7,-0.8) {{$\tilde \chi^0_1$}};
\node[black] at (6.75,0.8) {{$Z$}};
\node[black] at (8.3,0.7) {{$f$}};
\node[black] at (8.3,-0.2) {{$\bar f$}};

\end{scope}
\end{tikzpicture}
\caption{Radiative and three-body decay of the NLSP in the stealth NMSSM scenario.}
\label{fig:nmssm_extra_decays}
\end{figure}

Let us, however, go one step beyond  and look at the other possible final states of $\tilde \chi^0_2$ in the given scenario. Clearly, if phase space and couplings are large enough, the decay $\chi^0_2 \to S \tilde \chi^0_1$ will dominate over all others. Departing from this assumption, then three-body decays as well as radiative decays, discussed in sec.~\ref{sec:model-ind}, need to be taken into account. Both of which have a much larger phase space available and feature a different coupling structure.  
In Fig.~\ref{fig:nmssm_extra_decays} we depict the dominant diagrams for these new decay modes.

Let us investigate in which cases these modes are relevant. Quite obviously, for the tree-level decays to happen, a mixing between the bino and the singlino states is necessary -- which mainly proceeds via their higgsino admixtures. The latter is controlled by $\lambda$.
The coupling $\tilde \chi^0_2 - \tilde \chi^0_1 - h_S$ is governed by
\begin{align}
\lambda (Z_{\tilde \chi^0_2,\tilde H_d} Z_{\tilde \chi^0_1,\tilde H_u} + Z_{\tilde \chi^0_2,\tilde H_u} Z_{\tilde \chi^0_1,\tilde H_d})\,
\end{align}
where $Z_{\tilde \chi^0_i,\tilde H_j}$ is the $\tilde H_j$-admixture within the $i$-th neutralino.
This admixture must remain small in order to prevent direct decays $\tilde g \to j j \tilde \chi^0_1$ which would destroy the stealth setting due to a strong boost to $\tilde \chi^0_1$. Consequently, $\lambda$ must be small.

The coupling $\tilde \chi^0_2 - \tilde \chi^0_1 - Z$ is dominated by gauge couplings
\begin{align}
(g_2 \cos\theta_W + g_1 \sin\theta_W) (Z_{\tilde \chi^0_2,\tilde H_d} Z_{\tilde \chi^0_1,\tilde H_d} - Z_{\tilde \chi^0_2,\tilde H_u} Z_{\tilde \chi^0_1,\tilde H_u})\,.
\end{align}
So, while the two-body decay requires a $\lambda$ insertion in both the vertex and the neutralino admixture, for the three-body decay only the latter is needed.
Instead, for the loop decay (proceeding via charginos and a charged Higgs), no higgsino admixture is necessary, and there is only a single $\lambda$ dependence through the $\tilde \chi^\pm -H^\mp - \tilde \chi^0_1$ vertex.

In summary, small $\lambda$ is required for a stealth NMSSM scenario -- but 
 the smaller $\lambda$, the more important the otherwise sub-leading three-body and radiative decays become. The other main dependence of the decay channels comes through the mass of the higgsinos, $m_{\tilde H}$ -- and therefore $\mu$, which (i) enters the bino-higgsino as well as the singlino-higgsino mixture and (ii) determines the size of the three-body decay since higgsinos run in the loop of the radiative decay.

We are going to check now how large the effects of the new decay modes can become.
For the numerical evaluation we have used \SARAH \cite{Staub:2008uz,Staub:2009bi,Staub:2010jh,Staub:2012pb,Staub:2015kfa} to create a model-dependent code based on \SPheno \cite{Porod:2003um,Porod:2011nf,Staub:2017jnp}. The functionality of the automatic calculation of the one-loop radiative decays is described in Ref. \cite{Goodsell:2017pdq}, which we will make use of in the following. We have checked the benchmark points against {\tt HiggsBounds}\cite{Bechtle:2008jh, Bechtle:2011sb, Bechtle:2013gu, Bechtle:2013wla} to be in agreement with Higgs experimental searches.

\begin{figure}
\centering
\includegraphics[width=\linewidth]{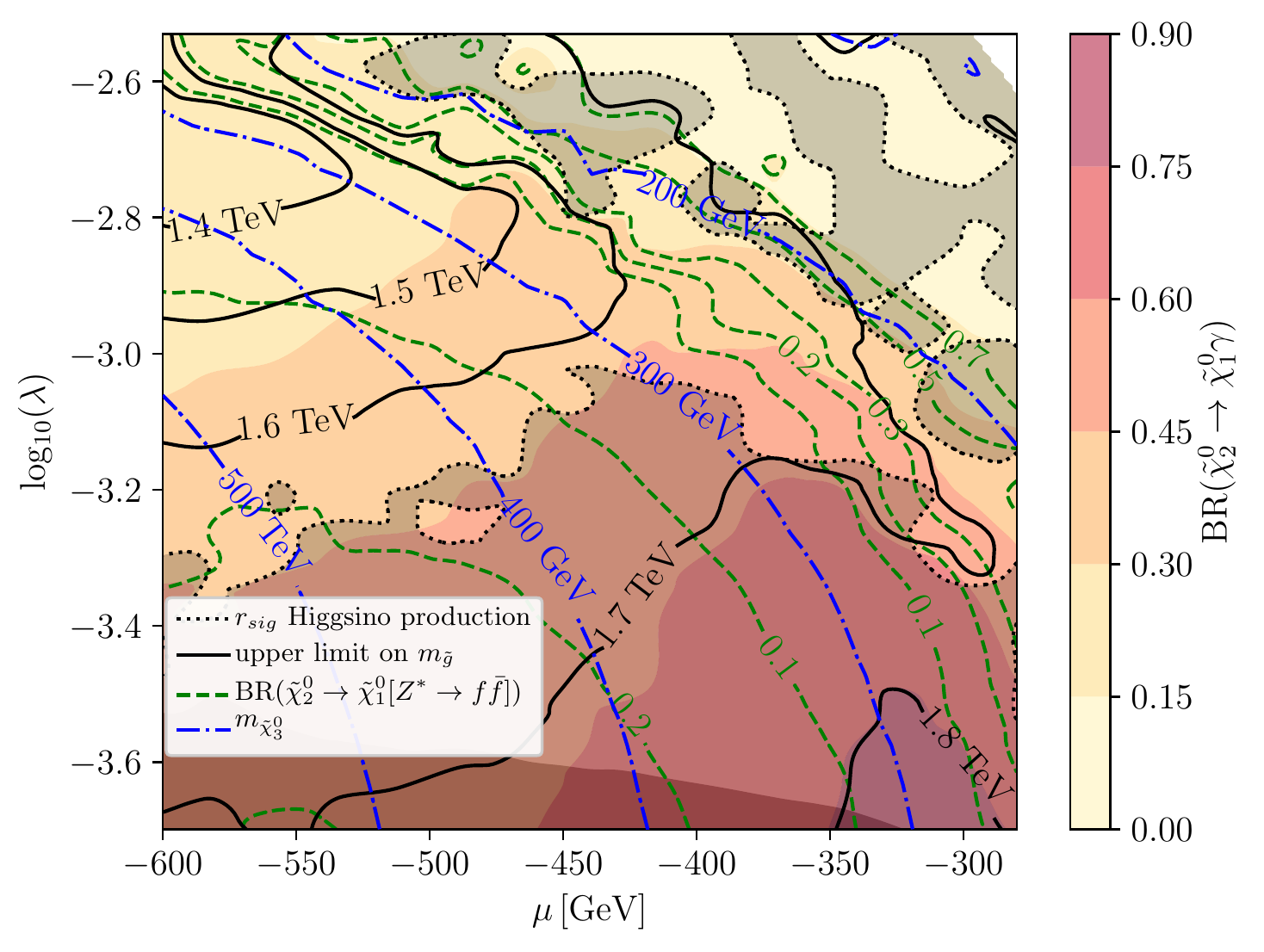}
\caption{Upper bounds on the gluino mass (black solid line) in the stealth NMSSM scenario as a function of $\mu$ and $\lambda$ while $\tan\beta=10$. The exclusion coming from Higgsino production is the light grey-shaded region.
The branching ratios of the second-to-lightest neutralino are shown in green dashed contours for the three-body decay and as a coloured background shading for the radiative decay. The mass of the third neutralino, $m_{\tilde{\chi}_3^0}$ is shown in blue dot-dashed contours.
We keep $\Delta m = m_{\tilde \chi^0_2} - m_{\tilde \chi^0_1} - m_S = 0.5\,$GeV while $m_{\tilde \chi^0_2} \simeq 89\,$GeV and  $m_{\tilde \chi^0_1} \simeq 5\,$GeV. 
The dark grey-shaded area at the bottom of the plot indicates the region where the total decay width of $\tilde \chi^0_2$ becomes smaller than $10^{-14}\,$GeV.
}
\label{fig:nmssmplot}
\end{figure}

In Fig.~\ref{fig:nmssmplot}, we plot the branching ratios of the three-body as well as the radiative decay as a function of $\mu$ and $\lambda$. Here we kept the mass gap of the two-body decay fixed at $m_{\tilde \chi^0_2}-m_{\tilde \chi^0_1}- m_{S} = 0.5\,$GeV. This was achieved by fitting this combination while adjusting  $v_S, M_1, L_S$ for each point in parameter space. 
As expected, in this region of small $\lambda$, the importance of the alternative decay modes is huge as they sum up to almost 100\,\% in regions of small $|\mu|$. When increasing $|\mu|$ towards larger values, the branching ratio of the radiative decay is reduced due to the increase in chargino mass. Throughout the plane, however, we find that the stealth decay mode is only sub-leading so that the signal at the LHC would indeed feature significant $\met$ and  photons, providing completely different prospects of discovery. Indeed, while the recasted LHC analyses of section~\ref{sec:model-ind} only exclude gluino masses of $\sim 1.4\,$TeV in the upper left part of the plot, the exclusion power reaches $m_{\tilde g} = 1.8\,$TeV in the lower right corner with small $\lambda$ and $|\mu|$.
We see when comparing the contour lines of the radiative decay and the gluino exclusion lines that for large $m_{\tilde g}$, the LHC exclusion power is dominated by the photon searches of the likes of Ref.~\cite{Aaboud:2018doq}, analogous to the corner of Fig.~\ref{fig:simpmodelsgeneral} where the photonic decay dominates.  For smaller gluino masses and therefore larger production cross sections, however, also the upper left region of Fig.~\ref{fig:nmssmplot} where the three-body decay dominates is covered by jets+$\met$ searches like Ref.~\cite{Aaboud:2017vwy}.

Finally, we also show the exclusion line from electroweakino searches due to the presence of light higgsinos as the light grey-shaded area surrounded by a black dotted line. By our choices of $\mu$ and $\lambda$, the higgsino mass varies from roughly 150\,GeV to 600\,GeV throughout the plot (blue dashed lines), leading to large differences in the production cross section of the higgsino. In combination with the varying branching ratios, we observe two areas where the higgsinos are excluded. The first one is in the top right corner where the second neutralino branching ratio into the $Z$ boson is enhanced and we have a third neutralino with masses equal or less than $m_{\tilde{\chi}_3^0}$ $\lesssim$ 200 GeV. The second interesting region is found in the bottom half of Fig.~\ref{fig:nmssmplot}. Here the excluded area tells us that the photonic search~\cite{Aaboud:2018doq} is sensitive to the larger branching ratio of the second lightest neutralino into photons. If we compare the range of masses and values of the branching ratio into photons for which the exclusion rate is higher we can see that they match with the ones obtained in Fig.~\ref{fig:simpmodelsgeneralhiggsino}. We find the best exclusion signal rates for masses between $m_{\tilde \chi^3_0}$=250 -- 600~GeV and branching ratios into photons greater than 70\%.

A comment about the total decay width $\Gamma$ of $\tilde \chi^0_2$ is in order. In the shown plane,  Fig.~\ref{fig:nmssmplot}, $\Gamma$ reaches down to $10^{-14}$\,GeV and slightly lower. The region where this happens is shaded in grey and is located at the bottom left of the plot.  
Because of this small width and the associated time-delayed decay, one might ask whether searches for non-pointing photons might be relevant. However,  while the readout (at ATLAS) features a time resolution of $\sim 70\,$ps \cite{1748-0221-5-09-P09003}, 
exclusion results are only presented for lifetimes of 250\,ps and more since for lower photon lifetimes, background rejection proves to be too difficult \cite{Aad:2014gfa}. Consequently, a conservative estimate is that lifetimes below $10^{-14}$\,GeV could indeed be resolved as non-pointing photons whereas above, they have to be tagged conventionally.



\section{Conclusion}
\label{sec:conclusion}

We have examined the possibility that additional decay channels contribute to otherwise stealth SUSY scenarios. These are constructed such that the LSP is very light while the phase space of the tree-level NLSP two-body decay is very small. Coloured production at the LHC then eventually leads to signals of several jets but almost no $\met$. Because of the reduced phase space, however, other suppressed decay channels of the NLSP, such as three-body and radiative, can also become relevant and even dominate. We have shown that already for small contributions to the branching fraction, these extra decays weaken the appealing features of stealth scenarios, meaning that the limits on the coloured sector become significantly stronger. Furthermore we have also compared the electroweakino production in stealth SUSY scenarios finding that the presence of the new decay rates could make them invisible. We have demonstrated this by recasting relevant LHC searches and calculating the limits on the gluino and higgsino masses depending on the NLSP branching ratios. 
We have then shown at the example of two realistic models that these extra decay modes are indeed relevant. In the St\"uckelberg extension of the MSSM, the stealth two-body decay is almost non-existent. In the NMSSM, we find regions of parameter space which are stealth directly next to regions which feature dominating three-body decays as well as dominating photonic final states. We have finally presented the gluino and higgsino mass limits in this NMSSM scenario and find differences of 400\,GeV and more between the different regions of parameter space.



%

\section*{Acknowledgements}
We thank Toby Opferkuch for interesting discussions and collaboration in the initial phase of this project. MEK is supported by the DFG Research Unit 2239 ``New Physics at the LHC''. VML acknowledges support of the BMBF under the project 05H15PDCAA. FS is supported by the ERC Recognition Award ERC-RA-0008 of the Helmholtz Association. \\

\bibliography{lit}

\begin{thebibliography}{56}%
\makeatletter
\providecommand \@ifxundefined [1]{%
 \@ifx{#1\undefined}
}%
\providecommand \@ifnum [1]{%
 \ifnum #1\expandafter \@firstoftwo
 \else \expandafter \@secondoftwo
 \fi
}%
\providecommand \@ifx [1]{%
 \ifx #1\expandafter \@firstoftwo
 \else \expandafter \@secondoftwo
 \fi
}%
\providecommand \natexlab [1]{#1}%
\providecommand \enquote  [1]{``#1''}%
\providecommand \bibnamefont  [1]{#1}%
\providecommand \bibfnamefont [1]{#1}%
\providecommand \citenamefont [1]{#1}%
\providecommand \href@noop [0]{\@secondoftwo}%
\providecommand \href [0]{\begingroup \@sanitize@url \@href}%
\providecommand \@href[1]{\@@startlink{#1}\@@href}%
\providecommand \@@href[1]{\endgroup#1\@@endlink}%
\providecommand \@sanitize@url [0]{\catcode `\\12\catcode `\$12\catcode
  `\&12\catcode `\#12\catcode `\^12\catcode `\_12\catcode `\%12\relax}%
\providecommand \@@startlink[1]{}%
\providecommand \@@endlink[0]{}%
\providecommand \url  [0]{\begingroup\@sanitize@url \@url }%
\providecommand \@url [1]{\endgroup\@href {#1}{\urlprefix }}%
\providecommand \urlprefix  [0]{URL }%
\providecommand \Eprint [0]{\href }%
\providecommand \doibase [0]{http://dx.doi.org/}%
\providecommand \selectlanguage [0]{\@gobble}%
\providecommand \bibinfo  [0]{\@secondoftwo}%
\providecommand \bibfield  [0]{\@secondoftwo}%
\providecommand \translation [1]{[#1]}%
\providecommand \BibitemOpen [0]{}%
\providecommand \bibitemStop [0]{}%
\providecommand \bibitemNoStop [0]{.\EOS\space}%
\providecommand \EOS [0]{\spacefactor3000\relax}%
\providecommand \BibitemShut  [1]{\csname bibitem#1\endcsname}%
\let\auto@bib@innerbib\@empty
\bibitem [{\citenamefont {Aad}\ \emph {et~al.}(2012)\citenamefont {Aad} \emph
  {et~al.}}]{Aad:2012tfa}%
  \BibitemOpen
  \bibfield  {author} {\bibinfo {author} {\bibfnamefont {G.}~\bibnamefont
  {Aad}} \emph {et~al.} (\bibinfo {collaboration} {ATLAS}),\ }\href {\doibase
  10.1016/j.physletb.2012.08.020} {\bibfield  {journal} {\bibinfo  {journal}
  {Phys. Lett.}\ }\textbf {\bibinfo {volume} {B716}},\ \bibinfo {pages} {1}
  (\bibinfo {year} {2012})},\ \Eprint {http://arxiv.org/abs/1207.7214}
  {arXiv:1207.7214 [hep-ex]} \BibitemShut {NoStop}%
\bibitem [{\citenamefont {Chatrchyan}\ \emph {et~al.}(2012)\citenamefont
  {Chatrchyan} \emph {et~al.}}]{Chatrchyan:2012xdj}%
  \BibitemOpen
  \bibfield  {author} {\bibinfo {author} {\bibfnamefont {S.}~\bibnamefont
  {Chatrchyan}} \emph {et~al.} (\bibinfo {collaboration} {CMS}),\ }\href
  {\doibase 10.1016/j.physletb.2012.08.021} {\bibfield  {journal} {\bibinfo
  {journal} {Phys. Lett.}\ }\textbf {\bibinfo {volume} {B716}},\ \bibinfo
  {pages} {30} (\bibinfo {year} {2012})},\ \Eprint
  {http://arxiv.org/abs/1207.7235} {arXiv:1207.7235 [hep-ex]} \BibitemShut
  {NoStop}%
\bibitem [{\citenamefont {Allanach}\ \emph {et~al.}(2002)\citenamefont
  {Allanach} \emph {et~al.}}]{Allanach:2002nj}%
  \BibitemOpen
  \bibfield  {author} {\bibinfo {author} {\bibfnamefont {B.~C.}\ \bibnamefont
  {Allanach}} \emph {et~al.},\ }\bibfield  {booktitle} {\emph {\bibinfo
  {booktitle} {{Proceedings, APS / DPF / DPB Summer Study on the Future of
  Particle Physics (Snowmass 2001), Snowmass, Colorado, 30 Jun - 21 Jul
  2001}}},\ }\href {\doibase 10.1007/s10052-002-0949-3} {\bibfield  {journal}
  {\bibinfo  {journal} {Eur. Phys. J.}\ }\textbf {\bibinfo {volume} {C25}},\
  \bibinfo {pages} {113} (\bibinfo {year} {2002})},\ \Eprint
  {http://arxiv.org/abs/hep-ph/0202233} {arXiv:hep-ph/0202233 [hep-ph]}
  \BibitemShut {NoStop}%
\bibitem [{\citenamefont {CMS}(2017)}]{CMS:2017kmd}%
  \BibitemOpen
  \bibfield  {author} {\bibinfo {author} {\bibnamefont {CMS}},\ }\href@noop {}
  {\bibfield  {journal} {\bibinfo  {journal} {CMS-PAS-SUS-16-036}\ } (\bibinfo
  {year} {2017})}\BibitemShut {NoStop}%
\bibitem [{\citenamefont {Aaboud}\ \emph
  {et~al.}(2018{\natexlab{a}})\citenamefont {Aaboud} \emph
  {et~al.}}]{Aaboud:2017vwy}%
  \BibitemOpen
  \bibfield  {author} {\bibinfo {author} {\bibfnamefont {M.}~\bibnamefont
  {Aaboud}} \emph {et~al.} (\bibinfo {collaboration} {ATLAS}),\ }\href
  {\doibase 10.1103/PhysRevD.97.112001} {\bibfield  {journal} {\bibinfo
  {journal} {Phys. Rev.}\ }\textbf {\bibinfo {volume} {D97}},\ \bibinfo {pages}
  {112001} (\bibinfo {year} {2018}{\natexlab{a}})},\ \Eprint
  {http://arxiv.org/abs/1712.02332} {arXiv:1712.02332 [hep-ex]} \BibitemShut
  {NoStop}%
\bibitem [{\citenamefont {Dercks}\ \emph
  {et~al.}(2017{\natexlab{a}})\citenamefont {Dercks}, \citenamefont {Dreiner},
  \citenamefont {Krauss}, \citenamefont {Opferkuch},\ and\ \citenamefont
  {Reinert}}]{Dercks:2017lfq}%
  \BibitemOpen
  \bibfield  {author} {\bibinfo {author} {\bibfnamefont {D.}~\bibnamefont
  {Dercks}}, \bibinfo {author} {\bibfnamefont {H.}~\bibnamefont {Dreiner}},
  \bibinfo {author} {\bibfnamefont {M.~E.}\ \bibnamefont {Krauss}}, \bibinfo
  {author} {\bibfnamefont {T.}~\bibnamefont {Opferkuch}}, \ and\ \bibinfo
  {author} {\bibfnamefont {A.}~\bibnamefont {Reinert}},\ }\href {\doibase
  10.1140/epjc/s10052-017-5414-4} {\bibfield  {journal} {\bibinfo  {journal}
  {Eur. Phys. J.}\ }\textbf {\bibinfo {volume} {C77}},\ \bibinfo {pages} {856}
  (\bibinfo {year} {2017}{\natexlab{a}})},\ \Eprint
  {http://arxiv.org/abs/1706.09418} {arXiv:1706.09418 [hep-ph]} \BibitemShut
  {NoStop}%
\bibitem [{\citenamefont {Hanussek}\ and\ \citenamefont
  {Kim}(2012)}]{Hanussek:2012eh}%
  \BibitemOpen
  \bibfield  {author} {\bibinfo {author} {\bibfnamefont {M.}~\bibnamefont
  {Hanussek}}\ and\ \bibinfo {author} {\bibfnamefont {J.~S.}\ \bibnamefont
  {Kim}},\ }\href {\doibase 10.1103/PhysRevD.85.115021} {\bibfield  {journal}
  {\bibinfo  {journal} {Phys. Rev.}\ }\textbf {\bibinfo {volume} {D85}},\
  \bibinfo {pages} {115021} (\bibinfo {year} {2012})},\ \Eprint
  {http://arxiv.org/abs/1205.0019} {arXiv:1205.0019 [hep-ph]} \BibitemShut
  {NoStop}%
\bibitem [{\citenamefont {Hanussek}\ and\ \citenamefont
  {Kim}(2013)}]{Hanussek:2012fc}%
  \BibitemOpen
  \bibfield  {author} {\bibinfo {author} {\bibfnamefont {M.}~\bibnamefont
  {Hanussek}}\ and\ \bibinfo {author} {\bibfnamefont {J.~S.}\ \bibnamefont
  {Kim}},\ }\href {\doibase 10.1103/PhysRevD.87.035002} {\bibfield  {journal}
  {\bibinfo  {journal} {Phys. Rev.}\ }\textbf {\bibinfo {volume} {D87}},\
  \bibinfo {pages} {035002} (\bibinfo {year} {2013})},\ \Eprint
  {http://arxiv.org/abs/1211.0725} {arXiv:1211.0725 [hep-ph]} \BibitemShut
  {NoStop}%
\bibitem [{\citenamefont {Carena}\ \emph {et~al.}(2008)\citenamefont {Carena},
  \citenamefont {Freitas},\ and\ \citenamefont {Wagner}}]{Carena:2008mj}%
  \BibitemOpen
  \bibfield  {author} {\bibinfo {author} {\bibfnamefont {M.}~\bibnamefont
  {Carena}}, \bibinfo {author} {\bibfnamefont {A.}~\bibnamefont {Freitas}}, \
  and\ \bibinfo {author} {\bibfnamefont {C.~E.~M.}\ \bibnamefont {Wagner}},\
  }\href {\doibase 10.1088/1126-6708/2008/10/109} {\bibfield  {journal}
  {\bibinfo  {journal} {JHEP}\ }\textbf {\bibinfo {volume} {10}},\ \bibinfo
  {pages} {109} (\bibinfo {year} {2008})},\ \Eprint
  {http://arxiv.org/abs/0808.2298} {arXiv:0808.2298 [hep-ph]} \BibitemShut
  {NoStop}%
\bibitem [{\citenamefont {Bornhauser}\ \emph {et~al.}(2011)\citenamefont
  {Bornhauser}, \citenamefont {Drees}, \citenamefont {Grab},\ and\
  \citenamefont {Kim}}]{Bornhauser:2010mw}%
  \BibitemOpen
  \bibfield  {author} {\bibinfo {author} {\bibfnamefont {S.}~\bibnamefont
  {Bornhauser}}, \bibinfo {author} {\bibfnamefont {M.}~\bibnamefont {Drees}},
  \bibinfo {author} {\bibfnamefont {S.}~\bibnamefont {Grab}}, \ and\ \bibinfo
  {author} {\bibfnamefont {J.~S.}\ \bibnamefont {Kim}},\ }\href {\doibase
  10.1103/PhysRevD.83.035008} {\bibfield  {journal} {\bibinfo  {journal} {Phys.
  Rev.}\ }\textbf {\bibinfo {volume} {D83}},\ \bibinfo {pages} {035008}
  (\bibinfo {year} {2011})},\ \Eprint {http://arxiv.org/abs/1011.5508}
  {arXiv:1011.5508 [hep-ph]} \BibitemShut {NoStop}%
\bibitem [{\citenamefont {Drees}\ \emph {et~al.}(2012)\citenamefont {Drees},
  \citenamefont {Hanussek},\ and\ \citenamefont {Kim}}]{Drees:2012dd}%
  \BibitemOpen
  \bibfield  {author} {\bibinfo {author} {\bibfnamefont {M.}~\bibnamefont
  {Drees}}, \bibinfo {author} {\bibfnamefont {M.}~\bibnamefont {Hanussek}}, \
  and\ \bibinfo {author} {\bibfnamefont {J.~S.}\ \bibnamefont {Kim}},\ }\href
  {\doibase 10.1103/PhysRevD.86.035024} {\bibfield  {journal} {\bibinfo
  {journal} {Phys. Rev.}\ }\textbf {\bibinfo {volume} {D86}},\ \bibinfo {pages}
  {035024} (\bibinfo {year} {2012})},\ \Eprint {http://arxiv.org/abs/1201.5714}
  {arXiv:1201.5714 [hep-ph]} \BibitemShut {NoStop}%
\bibitem [{\citenamefont {Aaboud}\ \emph
  {et~al.}(2018{\natexlab{b}})\citenamefont {Aaboud} \emph
  {et~al.}}]{Aaboud:2017phn}%
  \BibitemOpen
  \bibfield  {author} {\bibinfo {author} {\bibfnamefont {M.}~\bibnamefont
  {Aaboud}} \emph {et~al.} (\bibinfo {collaboration} {ATLAS}),\ }\href
  {\doibase 10.1007/JHEP01(2018)126} {\bibfield  {journal} {\bibinfo  {journal}
  {JHEP}\ }\textbf {\bibinfo {volume} {01}},\ \bibinfo {pages} {126} (\bibinfo
  {year} {2018}{\natexlab{b}})},\ \Eprint {http://arxiv.org/abs/1711.03301}
  {arXiv:1711.03301 [hep-ex]} \BibitemShut {NoStop}%
\bibitem [{\citenamefont {Boehm}\ \emph {et~al.}(2000)\citenamefont {Boehm},
  \citenamefont {Djouadi},\ and\ \citenamefont {Drees}}]{Boehm:1999bj}%
  \BibitemOpen
  \bibfield  {author} {\bibinfo {author} {\bibfnamefont {C.}~\bibnamefont
  {Boehm}}, \bibinfo {author} {\bibfnamefont {A.}~\bibnamefont {Djouadi}}, \
  and\ \bibinfo {author} {\bibfnamefont {M.}~\bibnamefont {Drees}},\ }\href
  {\doibase 10.1103/PhysRevD.62.035012} {\bibfield  {journal} {\bibinfo
  {journal} {Phys. Rev.}\ }\textbf {\bibinfo {volume} {D62}},\ \bibinfo {pages}
  {035012} (\bibinfo {year} {2000})},\ \Eprint
  {http://arxiv.org/abs/hep-ph/9911496} {arXiv:hep-ph/9911496 [hep-ph]}
  \BibitemShut {NoStop}%
\bibitem [{\citenamefont {Fan}\ \emph {et~al.}(2011)\citenamefont {Fan},
  \citenamefont {Reece},\ and\ \citenamefont {Ruderman}}]{Fan:2011yu}%
  \BibitemOpen
  \bibfield  {author} {\bibinfo {author} {\bibfnamefont {J.}~\bibnamefont
  {Fan}}, \bibinfo {author} {\bibfnamefont {M.}~\bibnamefont {Reece}}, \ and\
  \bibinfo {author} {\bibfnamefont {J.~T.}\ \bibnamefont {Ruderman}},\ }\href
  {\doibase 10.1007/JHEP11(2011)012} {\bibfield  {journal} {\bibinfo  {journal}
  {JHEP}\ }\textbf {\bibinfo {volume} {11}},\ \bibinfo {pages} {012} (\bibinfo
  {year} {2011})},\ \Eprint {http://arxiv.org/abs/1105.5135} {arXiv:1105.5135
  [hep-ph]} \BibitemShut {NoStop}%
\bibitem [{\citenamefont {Chatrchyan}\ \emph {et~al.}(2013)\citenamefont
  {Chatrchyan} \emph {et~al.}}]{CMS:2012un}%
  \BibitemOpen
  \bibfield  {author} {\bibinfo {author} {\bibfnamefont {S.}~\bibnamefont
  {Chatrchyan}} \emph {et~al.} (\bibinfo {collaboration} {CMS}),\ }\href
  {\doibase 10.1016/j.physletb.2012.12.055} {\bibfield  {journal} {\bibinfo
  {journal} {Phys. Lett.}\ }\textbf {\bibinfo {volume} {B719}},\ \bibinfo
  {pages} {42} (\bibinfo {year} {2013})},\ \Eprint
  {http://arxiv.org/abs/1210.2052} {arXiv:1210.2052 [hep-ex]} \BibitemShut
  {NoStop}%
\bibitem [{\citenamefont {Ellwanger}\ and\ \citenamefont
  {Teixeira}(2014)}]{Ellwanger:2014hia}%
  \BibitemOpen
  \bibfield  {author} {\bibinfo {author} {\bibfnamefont {U.}~\bibnamefont
  {Ellwanger}}\ and\ \bibinfo {author} {\bibfnamefont {A.~M.}\ \bibnamefont
  {Teixeira}},\ }\href {\doibase 10.1007/JHEP10(2014)113} {\bibfield  {journal}
  {\bibinfo  {journal} {JHEP}\ }\textbf {\bibinfo {volume} {10}},\ \bibinfo
  {pages} {113} (\bibinfo {year} {2014})},\ \Eprint
  {http://arxiv.org/abs/1406.7221} {arXiv:1406.7221 [hep-ph]} \BibitemShut
  {NoStop}%
\bibitem [{\citenamefont
  {collaboration}(2013)}]{TheATLAScollaboration:2013xia}%
  \BibitemOpen
  \bibfield  {author} {\bibinfo {author} {\bibfnamefont {T.~A.}\ \bibnamefont
  {collaboration}} (\bibinfo {collaboration} {ATLAS}),\ }\href@noop {} {\
  (\bibinfo {year} {2013})}\BibitemShut {NoStop}%
\bibitem [{\citenamefont {Khachatryan}\ \emph {et~al.}(2017)\citenamefont
  {Khachatryan} \emph {et~al.}}]{Khachatryan:2016xim}%
  \BibitemOpen
  \bibfield  {author} {\bibinfo {author} {\bibfnamefont {V.}~\bibnamefont
  {Khachatryan}} \emph {et~al.} (\bibinfo {collaboration} {CMS}),\ }\href
  {\doibase 10.1016/j.physletb.2017.01.073} {\bibfield  {journal} {\bibinfo
  {journal} {Phys. Lett.}\ }\textbf {\bibinfo {volume} {B770}},\ \bibinfo
  {pages} {257} (\bibinfo {year} {2017})},\ \Eprint
  {http://arxiv.org/abs/1608.01224} {arXiv:1608.01224 [hep-ex]} \BibitemShut
  {NoStop}%
\bibitem [{\citenamefont {Aaboud}\ \emph
  {et~al.}(2018{\natexlab{c}})\citenamefont {Aaboud} \emph
  {et~al.}}]{Aaboud:2018lpl}%
  \BibitemOpen
  \bibfield  {author} {\bibinfo {author} {\bibfnamefont {M.}~\bibnamefont
  {Aaboud}} \emph {et~al.} (\bibinfo {collaboration} {ATLAS}),\ }\href@noop {}
  {\  (\bibinfo {year} {2018}{\natexlab{c}})},\ \Eprint
  {http://arxiv.org/abs/1804.03568} {arXiv:1804.03568 [hep-ex]} \BibitemShut
  {NoStop}%
\bibitem [{\citenamefont {Sj{\"o}strand}\ \emph {et~al.}(2015)\citenamefont
  {Sj{\"o}strand}, \citenamefont {Ask}, \citenamefont {Christiansen},
  \citenamefont {Corke}, \citenamefont {Desai}, \citenamefont {Ilten},
  \citenamefont {Mrenna}, \citenamefont {Prestel}, \citenamefont {Rasmussen},\
  and\ \citenamefont {Skands}}]{Sjostrand:2014zea}%
  \BibitemOpen
  \bibfield  {author} {\bibinfo {author} {\bibfnamefont {T.}~\bibnamefont
  {Sj{\"o}strand}}, \bibinfo {author} {\bibfnamefont {S.}~\bibnamefont {Ask}},
  \bibinfo {author} {\bibfnamefont {J.~R.}\ \bibnamefont {Christiansen}},
  \bibinfo {author} {\bibfnamefont {R.}~\bibnamefont {Corke}}, \bibinfo
  {author} {\bibfnamefont {N.}~\bibnamefont {Desai}}, \bibinfo {author}
  {\bibfnamefont {P.}~\bibnamefont {Ilten}}, \bibinfo {author} {\bibfnamefont
  {S.}~\bibnamefont {Mrenna}}, \bibinfo {author} {\bibfnamefont
  {S.}~\bibnamefont {Prestel}}, \bibinfo {author} {\bibfnamefont {C.~O.}\
  \bibnamefont {Rasmussen}}, \ and\ \bibinfo {author} {\bibfnamefont {P.~Z.}\
  \bibnamefont {Skands}},\ }\href {\doibase 10.1016/j.cpc.2015.01.024}
  {\bibfield  {journal} {\bibinfo  {journal} {Comput. Phys. Commun.}\ }\textbf
  {\bibinfo {volume} {191}},\ \bibinfo {pages} {159} (\bibinfo {year}
  {2015})},\ \Eprint {http://arxiv.org/abs/1410.3012} {arXiv:1410.3012
  [hep-ph]} \BibitemShut {NoStop}%
\bibitem [{\citenamefont {Ball}\ \emph {et~al.}(2013)\citenamefont {Ball} \emph
  {et~al.}}]{Ball:2012cx}%
  \BibitemOpen
  \bibfield  {author} {\bibinfo {author} {\bibfnamefont {R.~D.}\ \bibnamefont
  {Ball}} \emph {et~al.},\ }\href {\doibase 10.1016/j.nuclphysb.2012.10.003}
  {\bibfield  {journal} {\bibinfo  {journal} {Nucl. Phys.}\ }\textbf {\bibinfo
  {volume} {B867}},\ \bibinfo {pages} {244} (\bibinfo {year} {2013})},\ \Eprint
  {http://arxiv.org/abs/1207.1303} {arXiv:1207.1303 [hep-ph]} \BibitemShut
  {NoStop}%
\bibitem [{\citenamefont {Beenakker}\ \emph {et~al.}(1997)\citenamefont
  {Beenakker}, \citenamefont {Hopker}, \citenamefont {Spira},\ and\
  \citenamefont {Zerwas}}]{Beenakker:1996ch}%
  \BibitemOpen
  \bibfield  {author} {\bibinfo {author} {\bibfnamefont {W.}~\bibnamefont
  {Beenakker}}, \bibinfo {author} {\bibfnamefont {R.}~\bibnamefont {Hopker}},
  \bibinfo {author} {\bibfnamefont {M.}~\bibnamefont {Spira}}, \ and\ \bibinfo
  {author} {\bibfnamefont {P.}~\bibnamefont {Zerwas}},\ }\href {\doibase
  10.1016/S0550-3213(97)80027-2} {\bibfield  {journal} {\bibinfo  {journal}
  {Nucl.Phys.}\ }\textbf {\bibinfo {volume} {B492}},\ \bibinfo {pages} {51}
  (\bibinfo {year} {1997})},\ \Eprint {http://arxiv.org/abs/hep-ph/9610490}
  {arXiv:hep-ph/9610490 [hep-ph]} \BibitemShut {NoStop}%
\bibitem [{\citenamefont {Beenakker}\ \emph {et~al.}(1998)\citenamefont
  {Beenakker}, \citenamefont {Kramer}, \citenamefont {Plehn}, \citenamefont
  {Spira},\ and\ \citenamefont {Zerwas}}]{Beenakker:1997ut}%
  \BibitemOpen
  \bibfield  {author} {\bibinfo {author} {\bibfnamefont {W.}~\bibnamefont
  {Beenakker}}, \bibinfo {author} {\bibfnamefont {M.}~\bibnamefont {Kramer}},
  \bibinfo {author} {\bibfnamefont {T.}~\bibnamefont {Plehn}}, \bibinfo
  {author} {\bibfnamefont {M.}~\bibnamefont {Spira}}, \ and\ \bibinfo {author}
  {\bibfnamefont {P.~M.}\ \bibnamefont {Zerwas}},\ }\href {\doibase
  10.1016/S0550-3213(98)00014-5} {\bibfield  {journal} {\bibinfo  {journal}
  {Nucl. Phys.}\ }\textbf {\bibinfo {volume} {B515}},\ \bibinfo {pages} {3}
  (\bibinfo {year} {1998})},\ \Eprint {http://arxiv.org/abs/hep-ph/9710451}
  {arXiv:hep-ph/9710451 [hep-ph]} \BibitemShut {NoStop}%
\bibitem [{\citenamefont {Kulesza}\ and\ \citenamefont
  {Motyka}(2009{\natexlab{a}})}]{Kulesza:2008jb}%
  \BibitemOpen
  \bibfield  {author} {\bibinfo {author} {\bibfnamefont {A.}~\bibnamefont
  {Kulesza}}\ and\ \bibinfo {author} {\bibfnamefont {L.}~\bibnamefont
  {Motyka}},\ }\href {\doibase 10.1103/PhysRevLett.102.111802} {\bibfield
  {journal} {\bibinfo  {journal} {Phys.Rev.Lett.}\ }\textbf {\bibinfo {volume}
  {102}},\ \bibinfo {pages} {111802} (\bibinfo {year} {2009}{\natexlab{a}})},\
  \Eprint {http://arxiv.org/abs/0807.2405} {arXiv:0807.2405 [hep-ph]}
  \BibitemShut {NoStop}%
\bibitem [{\citenamefont {Kulesza}\ and\ \citenamefont
  {Motyka}(2009{\natexlab{b}})}]{Kulesza:2009kq}%
  \BibitemOpen
  \bibfield  {author} {\bibinfo {author} {\bibfnamefont {A.}~\bibnamefont
  {Kulesza}}\ and\ \bibinfo {author} {\bibfnamefont {L.}~\bibnamefont
  {Motyka}},\ }\href {\doibase 10.1103/PhysRevD.80.095004} {\bibfield
  {journal} {\bibinfo  {journal} {Phys.Rev.}\ }\textbf {\bibinfo {volume}
  {D80}},\ \bibinfo {pages} {095004} (\bibinfo {year} {2009}{\natexlab{b}})},\
  \Eprint {http://arxiv.org/abs/0905.4749} {arXiv:0905.4749 [hep-ph]}
  \BibitemShut {NoStop}%
\bibitem [{\citenamefont {Beenakker}\ \emph {et~al.}(2009)\citenamefont
  {Beenakker}, \citenamefont {Brensing}, \citenamefont {Kramer}, \citenamefont
  {Kulesza}, \citenamefont {Laenen} \emph {et~al.}}]{Beenakker:2009ha}%
  \BibitemOpen
  \bibfield  {author} {\bibinfo {author} {\bibfnamefont {W.}~\bibnamefont
  {Beenakker}}, \bibinfo {author} {\bibfnamefont {S.}~\bibnamefont {Brensing}},
  \bibinfo {author} {\bibfnamefont {M.}~\bibnamefont {Kramer}}, \bibinfo
  {author} {\bibfnamefont {A.}~\bibnamefont {Kulesza}}, \bibinfo {author}
  {\bibfnamefont {E.}~\bibnamefont {Laenen}},  \emph {et~al.},\ }\href
  {\doibase 10.1088/1126-6708/2009/12/041} {\bibfield  {journal} {\bibinfo
  {journal} {JHEP}\ }\textbf {\bibinfo {volume} {0912}},\ \bibinfo {pages}
  {041} (\bibinfo {year} {2009})},\ \Eprint {http://arxiv.org/abs/0909.4418}
  {arXiv:0909.4418 [hep-ph]} \BibitemShut {NoStop}%
\bibitem [{\citenamefont {Beenakker}\ \emph {et~al.}(2010)\citenamefont
  {Beenakker}, \citenamefont {Brensing}, \citenamefont {Kramer}, \citenamefont
  {Kulesza}, \citenamefont {Laenen},\ and\ \citenamefont
  {Niessen}}]{Beenakker:2010nq}%
  \BibitemOpen
  \bibfield  {author} {\bibinfo {author} {\bibfnamefont {W.}~\bibnamefont
  {Beenakker}}, \bibinfo {author} {\bibfnamefont {S.}~\bibnamefont {Brensing}},
  \bibinfo {author} {\bibfnamefont {M.}~\bibnamefont {Kramer}}, \bibinfo
  {author} {\bibfnamefont {A.}~\bibnamefont {Kulesza}}, \bibinfo {author}
  {\bibfnamefont {E.}~\bibnamefont {Laenen}}, \ and\ \bibinfo {author}
  {\bibfnamefont {I.}~\bibnamefont {Niessen}},\ }\href {\doibase
  10.1007/JHEP08(2010)098} {\bibfield  {journal} {\bibinfo  {journal} {JHEP}\
  }\textbf {\bibinfo {volume} {08}},\ \bibinfo {pages} {098} (\bibinfo {year}
  {2010})},\ \Eprint {http://arxiv.org/abs/1006.4771} {arXiv:1006.4771
  [hep-ph]} \BibitemShut {NoStop}%
\bibitem [{\citenamefont {Beenakker}\ \emph {et~al.}(2011)\citenamefont
  {Beenakker}, \citenamefont {Brensing}, \citenamefont {Kramer}, \citenamefont
  {Kulesza}, \citenamefont {Laenen} \emph {et~al.}}]{Beenakker:2011fu}%
  \BibitemOpen
  \bibfield  {author} {\bibinfo {author} {\bibfnamefont {W.}~\bibnamefont
  {Beenakker}}, \bibinfo {author} {\bibfnamefont {S.}~\bibnamefont {Brensing}},
  \bibinfo {author} {\bibfnamefont {M.}~\bibnamefont {Kramer}}, \bibinfo
  {author} {\bibfnamefont {A.}~\bibnamefont {Kulesza}}, \bibinfo {author}
  {\bibfnamefont {E.}~\bibnamefont {Laenen}},  \emph {et~al.},\ }\href
  {\doibase 10.1142/S0217751X11053560} {\bibfield  {journal} {\bibinfo
  {journal} {Int.J.Mod.Phys.}\ }\textbf {\bibinfo {volume} {A26}},\ \bibinfo
  {pages} {2637} (\bibinfo {year} {2011})},\ \Eprint
  {http://arxiv.org/abs/1105.1110} {arXiv:1105.1110 [hep-ph]} \BibitemShut
  {NoStop}%
\bibitem [{\citenamefont {Drees}\ \emph {et~al.}(2015)\citenamefont {Drees},
  \citenamefont {Dreiner}, \citenamefont {Schmeier}, \citenamefont
  {Tattersall},\ and\ \citenamefont {Kim}}]{Drees:2013wra}%
  \BibitemOpen
  \bibfield  {author} {\bibinfo {author} {\bibfnamefont {M.}~\bibnamefont
  {Drees}}, \bibinfo {author} {\bibfnamefont {H.}~\bibnamefont {Dreiner}},
  \bibinfo {author} {\bibfnamefont {D.}~\bibnamefont {Schmeier}}, \bibinfo
  {author} {\bibfnamefont {J.}~\bibnamefont {Tattersall}}, \ and\ \bibinfo
  {author} {\bibfnamefont {J.~S.}\ \bibnamefont {Kim}},\ }\href {\doibase
  10.1016/j.cpc.2014.10.018} {\bibfield  {journal} {\bibinfo  {journal}
  {Comput. Phys. Commun.}\ }\textbf {\bibinfo {volume} {187}},\ \bibinfo
  {pages} {227} (\bibinfo {year} {2015})},\ \Eprint
  {http://arxiv.org/abs/1312.2591} {arXiv:1312.2591 [hep-ph]} \BibitemShut
  {NoStop}%
\bibitem [{\citenamefont {Kim}\ \emph {et~al.}(2015)\citenamefont {Kim},
  \citenamefont {Schmeier}, \citenamefont {Tattersall},\ and\ \citenamefont
  {Rolbiecki}}]{Kim:2015wza}%
  \BibitemOpen
  \bibfield  {author} {\bibinfo {author} {\bibfnamefont {J.~S.}\ \bibnamefont
  {Kim}}, \bibinfo {author} {\bibfnamefont {D.}~\bibnamefont {Schmeier}},
  \bibinfo {author} {\bibfnamefont {J.}~\bibnamefont {Tattersall}}, \ and\
  \bibinfo {author} {\bibfnamefont {K.}~\bibnamefont {Rolbiecki}},\ }\href
  {\doibase 10.1016/j.cpc.2015.06.002} {\bibfield  {journal} {\bibinfo
  {journal} {Comput. Phys. Commun.}\ }\textbf {\bibinfo {volume} {196}},\
  \bibinfo {pages} {535} (\bibinfo {year} {2015})},\ \Eprint
  {http://arxiv.org/abs/1503.01123} {arXiv:1503.01123 [hep-ph]} \BibitemShut
  {NoStop}%
\bibitem [{\citenamefont {Dercks}\ \emph
  {et~al.}(2017{\natexlab{b}})\citenamefont {Dercks}, \citenamefont {Desai},
  \citenamefont {Kim}, \citenamefont {Rolbiecki}, \citenamefont {Tattersall},\
  and\ \citenamefont {Weber}}]{Dercks:2016npn}%
  \BibitemOpen
  \bibfield  {author} {\bibinfo {author} {\bibfnamefont {D.}~\bibnamefont
  {Dercks}}, \bibinfo {author} {\bibfnamefont {N.}~\bibnamefont {Desai}},
  \bibinfo {author} {\bibfnamefont {J.~S.}\ \bibnamefont {Kim}}, \bibinfo
  {author} {\bibfnamefont {K.}~\bibnamefont {Rolbiecki}}, \bibinfo {author}
  {\bibfnamefont {J.}~\bibnamefont {Tattersall}}, \ and\ \bibinfo {author}
  {\bibfnamefont {T.}~\bibnamefont {Weber}},\ }\href {\doibase
  10.1016/j.cpc.2017.08.021} {\bibfield  {journal} {\bibinfo  {journal}
  {Comput. Phys. Commun.}\ }\textbf {\bibinfo {volume} {221}},\ \bibinfo
  {pages} {383} (\bibinfo {year} {2017}{\natexlab{b}})},\ \Eprint
  {http://arxiv.org/abs/1611.09856} {arXiv:1611.09856 [hep-ph]} \BibitemShut
  {NoStop}%
\bibitem [{\citenamefont {de~Favereau}\ \emph {et~al.}(2014)\citenamefont
  {de~Favereau}, \citenamefont {Delaere}, \citenamefont {Demin}, \citenamefont
  {Giammanco}, \citenamefont {Lema\^{i}tre}, \citenamefont {Mertens},\ and\
  \citenamefont {Selvaggi}}]{deFavereau:2013fsa}%
  \BibitemOpen
  \bibfield  {author} {\bibinfo {author} {\bibfnamefont {J.}~\bibnamefont
  {de~Favereau}}, \bibinfo {author} {\bibfnamefont {C.}~\bibnamefont
  {Delaere}}, \bibinfo {author} {\bibfnamefont {P.}~\bibnamefont {Demin}},
  \bibinfo {author} {\bibfnamefont {A.}~\bibnamefont {Giammanco}}, \bibinfo
  {author} {\bibfnamefont {V.}~\bibnamefont {Lema\^{i}tre}}, \bibinfo {author}
  {\bibfnamefont {A.}~\bibnamefont {Mertens}}, \ and\ \bibinfo {author}
  {\bibfnamefont {M.}~\bibnamefont {Selvaggi}} (\bibinfo {collaboration}
  {DELPHES 3}),\ }\href {\doibase 10.1007/JHEP02(2014)057} {\bibfield
  {journal} {\bibinfo  {journal} {JHEP}\ }\textbf {\bibinfo {volume} {02}},\
  \bibinfo {pages} {057} (\bibinfo {year} {2014})},\ \Eprint
  {http://arxiv.org/abs/1307.6346} {arXiv:1307.6346 [hep-ex]} \BibitemShut
  {NoStop}%
\bibitem [{\citenamefont {Cacciari}\ \emph {et~al.}(2012)\citenamefont
  {Cacciari}, \citenamefont {Salam},\ and\ \citenamefont
  {Soyez}}]{Cacciari:2011ma}%
  \BibitemOpen
  \bibfield  {author} {\bibinfo {author} {\bibfnamefont {M.}~\bibnamefont
  {Cacciari}}, \bibinfo {author} {\bibfnamefont {G.~P.}\ \bibnamefont {Salam}},
  \ and\ \bibinfo {author} {\bibfnamefont {G.}~\bibnamefont {Soyez}},\ }\href
  {\doibase 10.1140/epjc/s10052-012-1896-2} {\bibfield  {journal} {\bibinfo
  {journal} {Eur. Phys. J.}\ }\textbf {\bibinfo {volume} {C72}},\ \bibinfo
  {pages} {1896} (\bibinfo {year} {2012})},\ \Eprint
  {http://arxiv.org/abs/1111.6097} {arXiv:1111.6097 [hep-ph]} \BibitemShut
  {NoStop}%
\bibitem [{\citenamefont {Cacciari}\ and\ \citenamefont
  {Salam}(2006)}]{Cacciari:2005hq}%
  \BibitemOpen
  \bibfield  {author} {\bibinfo {author} {\bibfnamefont {M.}~\bibnamefont
  {Cacciari}}\ and\ \bibinfo {author} {\bibfnamefont {G.~P.}\ \bibnamefont
  {Salam}},\ }\href {\doibase 10.1016/j.physletb.2006.08.037} {\bibfield
  {journal} {\bibinfo  {journal} {Phys. Lett.}\ }\textbf {\bibinfo {volume}
  {B641}},\ \bibinfo {pages} {57} (\bibinfo {year} {2006})},\ \Eprint
  {http://arxiv.org/abs/hep-ph/0512210} {arXiv:hep-ph/0512210 [hep-ph]}
  \BibitemShut {NoStop}%
\bibitem [{\citenamefont {Aaboud}\ \emph
  {et~al.}(2018{\natexlab{d}})\citenamefont {Aaboud} \emph
  {et~al.}}]{Aaboud:2018doq}%
  \BibitemOpen
  \bibfield  {author} {\bibinfo {author} {\bibfnamefont {M.}~\bibnamefont
  {Aaboud}} \emph {et~al.} (\bibinfo {collaboration} {ATLAS}),\ }\href
  {\doibase 10.1103/PhysRevD.97.092006} {\bibfield  {journal} {\bibinfo
  {journal} {Phys. Rev.}\ }\textbf {\bibinfo {volume} {D97}},\ \bibinfo {pages}
  {092006} (\bibinfo {year} {2018}{\natexlab{d}})},\ \Eprint
  {http://arxiv.org/abs/1802.03158} {arXiv:1802.03158 [hep-ex]} \BibitemShut
  {NoStop}%
\bibitem [{\citenamefont {Aaboud}\ \emph {et~al.}(2016)\citenamefont {Aaboud}
  \emph {et~al.}}]{ATLASCollaboration:2016wlb}%
  \BibitemOpen
  \bibfield  {author} {\bibinfo {author} {\bibfnamefont {M.}~\bibnamefont
  {Aaboud}} \emph {et~al.} (\bibinfo {collaboration} {ATLAS}),\ }\href
  {\doibase 10.1140/epjc/s10052-016-4344-x} {\bibfield  {journal} {\bibinfo
  {journal} {Eur. Phys. J.}\ }\textbf {\bibinfo {volume} {C76}},\ \bibinfo
  {pages} {517} (\bibinfo {year} {2016})},\ \Eprint
  {http://arxiv.org/abs/1606.09150} {arXiv:1606.09150 [hep-ex]} \BibitemShut
  {NoStop}%
\bibitem [{\citenamefont {Sirunyan}\ \emph {et~al.}(2018)\citenamefont
  {Sirunyan} \emph {et~al.}}]{Sirunyan:2017lae}%
  \BibitemOpen
  \bibfield  {author} {\bibinfo {author} {\bibfnamefont {A.~M.}\ \bibnamefont
  {Sirunyan}} \emph {et~al.} (\bibinfo {collaboration} {CMS}),\ }\href
  {\doibase 10.1007/JHEP03(2018)166} {\bibfield  {journal} {\bibinfo  {journal}
  {JHEP}\ }\textbf {\bibinfo {volume} {03}},\ \bibinfo {pages} {166} (\bibinfo
  {year} {2018})},\ \Eprint {http://arxiv.org/abs/1709.05406} {arXiv:1709.05406
  [hep-ex]} \BibitemShut {NoStop}%
\bibitem [{\citenamefont {Kors}\ and\ \citenamefont
  {Nath}(2004)}]{Kors:2004ri}%
  \BibitemOpen
  \bibfield  {author} {\bibinfo {author} {\bibfnamefont {B.}~\bibnamefont
  {Kors}}\ and\ \bibinfo {author} {\bibfnamefont {P.}~\bibnamefont {Nath}},\
  }\href {\doibase 10.1088/1126-6708/2004/12/005} {\bibfield  {journal}
  {\bibinfo  {journal} {JHEP}\ }\textbf {\bibinfo {volume} {12}},\ \bibinfo
  {pages} {005} (\bibinfo {year} {2004})},\ \Eprint
  {http://arxiv.org/abs/hep-ph/0406167} {arXiv:hep-ph/0406167 [hep-ph]}
  \BibitemShut {NoStop}%
\bibitem [{\citenamefont {Feldman}\ \emph {et~al.}(2006)\citenamefont
  {Feldman}, \citenamefont {Liu},\ and\ \citenamefont {Nath}}]{Feldman:2006wb}%
  \BibitemOpen
  \bibfield  {author} {\bibinfo {author} {\bibfnamefont {D.}~\bibnamefont
  {Feldman}}, \bibinfo {author} {\bibfnamefont {Z.}~\bibnamefont {Liu}}, \ and\
  \bibinfo {author} {\bibfnamefont {P.}~\bibnamefont {Nath}},\ }\href {\doibase
  10.1088/1126-6708/2006/11/007} {\bibfield  {journal} {\bibinfo  {journal}
  {JHEP}\ }\textbf {\bibinfo {volume} {11}},\ \bibinfo {pages} {007} (\bibinfo
  {year} {2006})},\ \Eprint {http://arxiv.org/abs/hep-ph/0606294}
  {arXiv:hep-ph/0606294 [hep-ph]} \BibitemShut {NoStop}%
\bibitem [{\citenamefont {Ellwanger}\ and\ \citenamefont
  {Teixeira}(2015)}]{Ellwanger:2014hca}%
  \BibitemOpen
  \bibfield  {author} {\bibinfo {author} {\bibfnamefont {U.}~\bibnamefont
  {Ellwanger}}\ and\ \bibinfo {author} {\bibfnamefont {A.~M.}\ \bibnamefont
  {Teixeira}},\ }\href {\doibase 10.1007/JHEP04(2015)172} {\bibfield  {journal}
  {\bibinfo  {journal} {JHEP}\ }\textbf {\bibinfo {volume} {04}},\ \bibinfo
  {pages} {172} (\bibinfo {year} {2015})},\ \Eprint
  {http://arxiv.org/abs/1412.6394} {arXiv:1412.6394 [hep-ph]} \BibitemShut
  {NoStop}%
\bibitem [{\citenamefont {Titterton}\ \emph {et~al.}(2018)\citenamefont
  {Titterton}, \citenamefont {Ellwanger}, \citenamefont {Flaecher},
  \citenamefont {Moretti},\ and\ \citenamefont
  {Shepherd-Themistocleous}}]{Titterton:2018pba}%
  \BibitemOpen
  \bibfield  {author} {\bibinfo {author} {\bibfnamefont {A.}~\bibnamefont
  {Titterton}}, \bibinfo {author} {\bibfnamefont {U.}~\bibnamefont
  {Ellwanger}}, \bibinfo {author} {\bibfnamefont {H.~U.}\ \bibnamefont
  {Flaecher}}, \bibinfo {author} {\bibfnamefont {S.}~\bibnamefont {Moretti}}, \
  and\ \bibinfo {author} {\bibfnamefont {C.~H.}\ \bibnamefont
  {Shepherd-Themistocleous}},\ }\href@noop {} {\  (\bibinfo {year} {2018})},\
  \Eprint {http://arxiv.org/abs/1807.10672} {arXiv:1807.10672 [hep-ph]}
  \BibitemShut {NoStop}%
\bibitem [{\citenamefont {Staub}(2008)}]{Staub:2008uz}%
  \BibitemOpen
  \bibfield  {author} {\bibinfo {author} {\bibfnamefont {F.}~\bibnamefont
  {Staub}},\ }\href@noop {} {\  (\bibinfo {year} {2008})},\ \Eprint
  {http://arxiv.org/abs/0806.0538} {arXiv:0806.0538 [hep-ph]} \BibitemShut
  {NoStop}%
\bibitem [{\citenamefont {Staub}(2010)}]{Staub:2009bi}%
  \BibitemOpen
  \bibfield  {author} {\bibinfo {author} {\bibfnamefont {F.}~\bibnamefont
  {Staub}},\ }\href {\doibase 10.1016/j.cpc.2010.01.011} {\bibfield  {journal}
  {\bibinfo  {journal} {Comput.Phys.Commun.}\ }\textbf {\bibinfo {volume}
  {181}},\ \bibinfo {pages} {1077} (\bibinfo {year} {2010})},\ \Eprint
  {http://arxiv.org/abs/0909.2863} {arXiv:0909.2863 [hep-ph]} \BibitemShut
  {NoStop}%
\bibitem [{\citenamefont {Staub}(2011)}]{Staub:2010jh}%
  \BibitemOpen
  \bibfield  {author} {\bibinfo {author} {\bibfnamefont {F.}~\bibnamefont
  {Staub}},\ }\href {\doibase 10.1016/j.cpc.2010.11.030} {\bibfield  {journal}
  {\bibinfo  {journal} {Comput.Phys.Commun.}\ }\textbf {\bibinfo {volume}
  {182}},\ \bibinfo {pages} {808} (\bibinfo {year} {2011})},\ \Eprint
  {http://arxiv.org/abs/1002.0840} {arXiv:1002.0840 [hep-ph]} \BibitemShut
  {NoStop}%
\bibitem [{\citenamefont {Staub}(2013)}]{Staub:2012pb}%
  \BibitemOpen
  \bibfield  {author} {\bibinfo {author} {\bibfnamefont {F.}~\bibnamefont
  {Staub}},\ }\href {\doibase 10.1016/j.cpc.2013.02.019} {\bibfield  {journal}
  {\bibinfo  {journal} {Comput. Phys. Commun.}\ }\textbf {\bibinfo {volume}
  {184}},\ \bibinfo {pages} {1792} (\bibinfo {year} {2013})},\ \Eprint
  {http://arxiv.org/abs/1207.0906} {arXiv:1207.0906 [hep-ph]} \BibitemShut
  {NoStop}%
\bibitem [{\citenamefont {Staub}(2015)}]{Staub:2015kfa}%
  \BibitemOpen
  \bibfield  {author} {\bibinfo {author} {\bibfnamefont {F.}~\bibnamefont
  {Staub}},\ }\href {\doibase 10.1155/2015/840780} {\bibfield  {journal}
  {\bibinfo  {journal} {Adv. High Energy Phys.}\ }\textbf {\bibinfo {volume}
  {2015}},\ \bibinfo {pages} {840780} (\bibinfo {year} {2015})},\ \Eprint
  {http://arxiv.org/abs/1503.04200} {arXiv:1503.04200 [hep-ph]} \BibitemShut
  {NoStop}%
\bibitem [{\citenamefont {Porod}(2003)}]{Porod:2003um}%
  \BibitemOpen
  \bibfield  {author} {\bibinfo {author} {\bibfnamefont {W.}~\bibnamefont
  {Porod}},\ }\href {\doibase 10.1016/S0010-4655(03)00222-4} {\bibfield
  {journal} {\bibinfo  {journal} {Comput.Phys.Commun.}\ }\textbf {\bibinfo
  {volume} {153}},\ \bibinfo {pages} {275} (\bibinfo {year} {2003})},\ \Eprint
  {http://arxiv.org/abs/hep-ph/0301101} {arXiv:hep-ph/0301101 [hep-ph]}
  \BibitemShut {NoStop}%
\bibitem [{\citenamefont {Porod}\ and\ \citenamefont
  {Staub}(2011)}]{Porod:2011nf}%
  \BibitemOpen
  \bibfield  {author} {\bibinfo {author} {\bibfnamefont {W.}~\bibnamefont
  {Porod}}\ and\ \bibinfo {author} {\bibfnamefont {F.}~\bibnamefont {Staub}},\
  }\href@noop {} {\  (\bibinfo {year} {2011})},\ \Eprint
  {http://arxiv.org/abs/1104.1573} {arXiv:1104.1573 [hep-ph]} \BibitemShut
  {NoStop}%
\bibitem [{\citenamefont {Staub}\ and\ \citenamefont
  {Porod}(2017)}]{Staub:2017jnp}%
  \BibitemOpen
  \bibfield  {author} {\bibinfo {author} {\bibfnamefont {F.}~\bibnamefont
  {Staub}}\ and\ \bibinfo {author} {\bibfnamefont {W.}~\bibnamefont {Porod}},\
  }\href {\doibase 10.1140/epjc/s10052-017-4893-7} {\bibfield  {journal}
  {\bibinfo  {journal} {Eur. Phys. J.}\ }\textbf {\bibinfo {volume} {C77}},\
  \bibinfo {pages} {338} (\bibinfo {year} {2017})},\ \Eprint
  {http://arxiv.org/abs/1703.03267} {arXiv:1703.03267 [hep-ph]} \BibitemShut
  {NoStop}%
\bibitem [{\citenamefont {Goodsell}\ \emph {et~al.}(2017)\citenamefont
  {Goodsell}, \citenamefont {Liebler},\ and\ \citenamefont
  {Staub}}]{Goodsell:2017pdq}%
  \BibitemOpen
  \bibfield  {author} {\bibinfo {author} {\bibfnamefont {M.~D.}\ \bibnamefont
  {Goodsell}}, \bibinfo {author} {\bibfnamefont {S.}~\bibnamefont {Liebler}}, \
  and\ \bibinfo {author} {\bibfnamefont {F.}~\bibnamefont {Staub}},\ }\href
  {\doibase 10.1140/epjc/s10052-017-5259-x} {\bibfield  {journal} {\bibinfo
  {journal} {Eur. Phys. J.}\ }\textbf {\bibinfo {volume} {C77}},\ \bibinfo
  {pages} {758} (\bibinfo {year} {2017})},\ \Eprint
  {http://arxiv.org/abs/1703.09237} {arXiv:1703.09237 [hep-ph]} \BibitemShut
  {NoStop}%
\bibitem [{\citenamefont {Bechtle}\ \emph {et~al.}(2010)\citenamefont
  {Bechtle}, \citenamefont {Brein}, \citenamefont {Heinemeyer}, \citenamefont
  {Weiglein},\ and\ \citenamefont {Williams}}]{Bechtle:2008jh}%
  \BibitemOpen
  \bibfield  {author} {\bibinfo {author} {\bibfnamefont {P.}~\bibnamefont
  {Bechtle}}, \bibinfo {author} {\bibfnamefont {O.}~\bibnamefont {Brein}},
  \bibinfo {author} {\bibfnamefont {S.}~\bibnamefont {Heinemeyer}}, \bibinfo
  {author} {\bibfnamefont {G.}~\bibnamefont {Weiglein}}, \ and\ \bibinfo
  {author} {\bibfnamefont {K.~E.}\ \bibnamefont {Williams}},\ }\href {\doibase
  10.1016/j.cpc.2009.09.003} {\bibfield  {journal} {\bibinfo  {journal}
  {Comput. Phys. Commun.}\ }\textbf {\bibinfo {volume} {181}},\ \bibinfo
  {pages} {138} (\bibinfo {year} {2010})},\ \Eprint
  {http://arxiv.org/abs/0811.4169} {arXiv:0811.4169 [hep-ph]} \BibitemShut
  {NoStop}%
\bibitem [{\citenamefont {Bechtle}\ \emph {et~al.}(2011)\citenamefont
  {Bechtle}, \citenamefont {Brein}, \citenamefont {Heinemeyer}, \citenamefont
  {Weiglein},\ and\ \citenamefont {Williams}}]{Bechtle:2011sb}%
  \BibitemOpen
  \bibfield  {author} {\bibinfo {author} {\bibfnamefont {P.}~\bibnamefont
  {Bechtle}}, \bibinfo {author} {\bibfnamefont {O.}~\bibnamefont {Brein}},
  \bibinfo {author} {\bibfnamefont {S.}~\bibnamefont {Heinemeyer}}, \bibinfo
  {author} {\bibfnamefont {G.}~\bibnamefont {Weiglein}}, \ and\ \bibinfo
  {author} {\bibfnamefont {K.~E.}\ \bibnamefont {Williams}},\ }\href {\doibase
  10.1016/j.cpc.2011.07.015} {\bibfield  {journal} {\bibinfo  {journal}
  {Comput. Phys. Commun.}\ }\textbf {\bibinfo {volume} {182}},\ \bibinfo
  {pages} {2605} (\bibinfo {year} {2011})},\ \Eprint
  {http://arxiv.org/abs/1102.1898} {arXiv:1102.1898 [hep-ph]} \BibitemShut
  {NoStop}%
\bibitem [{\citenamefont {Bechtle}\ \emph {et~al.}(2012)\citenamefont
  {Bechtle}, \citenamefont {Brein}, \citenamefont {Heinemeyer}, \citenamefont
  {Stal}, \citenamefont {Stefaniak}, \citenamefont {Weiglein},\ and\
  \citenamefont {Williams}}]{Bechtle:2013gu}%
  \BibitemOpen
  \bibfield  {author} {\bibinfo {author} {\bibfnamefont {P.}~\bibnamefont
  {Bechtle}}, \bibinfo {author} {\bibfnamefont {O.}~\bibnamefont {Brein}},
  \bibinfo {author} {\bibfnamefont {S.}~\bibnamefont {Heinemeyer}}, \bibinfo
  {author} {\bibfnamefont {O.}~\bibnamefont {Stal}}, \bibinfo {author}
  {\bibfnamefont {T.}~\bibnamefont {Stefaniak}}, \bibinfo {author}
  {\bibfnamefont {G.}~\bibnamefont {Weiglein}}, \ and\ \bibinfo {author}
  {\bibfnamefont {K.}~\bibnamefont {Williams}},\ }\bibfield  {booktitle} {\emph
  {\bibinfo {booktitle} {{Proceedings, 4th International Workshop on Prospects
  for Charged Higgs Discovery at Colliders (CHARGED 2012): Uppsala, Sweden,
  October 8-11, 2012}}},\ }\href {\doibase 10.22323/1.156.0024} {\bibfield
  {journal} {\bibinfo  {journal} {PoS}\ }\textbf {\bibinfo {volume}
  {CHARGED2012}},\ \bibinfo {pages} {024} (\bibinfo {year} {2012})},\ \Eprint
  {http://arxiv.org/abs/1301.2345} {arXiv:1301.2345 [hep-ph]} \BibitemShut
  {NoStop}%
\bibitem [{\citenamefont {Bechtle}\ \emph {et~al.}(2014)\citenamefont
  {Bechtle}, \citenamefont {Brein}, \citenamefont {Heinemeyer}, \citenamefont
  {Stål}, \citenamefont {Stefaniak}, \citenamefont {Weiglein},\ and\
  \citenamefont {Williams}}]{Bechtle:2013wla}%
  \BibitemOpen
  \bibfield  {author} {\bibinfo {author} {\bibfnamefont {P.}~\bibnamefont
  {Bechtle}}, \bibinfo {author} {\bibfnamefont {O.}~\bibnamefont {Brein}},
  \bibinfo {author} {\bibfnamefont {S.}~\bibnamefont {Heinemeyer}}, \bibinfo
  {author} {\bibfnamefont {O.}~\bibnamefont {Stål}}, \bibinfo {author}
  {\bibfnamefont {T.}~\bibnamefont {Stefaniak}}, \bibinfo {author}
  {\bibfnamefont {G.}~\bibnamefont {Weiglein}}, \ and\ \bibinfo {author}
  {\bibfnamefont {K.~E.}\ \bibnamefont {Williams}},\ }\href {\doibase
  10.1140/epjc/s10052-013-2693-2} {\bibfield  {journal} {\bibinfo  {journal}
  {Eur. Phys. J.}\ }\textbf {\bibinfo {volume} {C74}},\ \bibinfo {pages} {2693}
  (\bibinfo {year} {2014})},\ \Eprint {http://arxiv.org/abs/1311.0055}
  {arXiv:1311.0055 [hep-ph]} \BibitemShut {NoStop}%
\bibitem [{\citenamefont {Abreu}\ \emph {et~al.}(2010)\citenamefont {Abreu}
  \emph {et~al.}}]{1748-0221-5-09-P09003}%
  \BibitemOpen
  \bibfield  {author} {\bibinfo {author} {\bibfnamefont {H.}~\bibnamefont
  {Abreu}} \emph {et~al.},\ }\href
  {http://stacks.iop.org/1748-0221/5/i=09/a=P09003} {\bibfield  {journal}
  {\bibinfo  {journal} {Journal of Instrumentation}\ }\textbf {\bibinfo
  {volume} {5}},\ \bibinfo {pages} {P09003} (\bibinfo {year}
  {2010})}\BibitemShut {NoStop}%
\bibitem [{\citenamefont {Aad}\ \emph {et~al.}(2014)\citenamefont {Aad} \emph
  {et~al.}}]{Aad:2014gfa}%
  \BibitemOpen
  \bibfield  {author} {\bibinfo {author} {\bibfnamefont {G.}~\bibnamefont
  {Aad}} \emph {et~al.} (\bibinfo {collaboration} {ATLAS}),\ }\href {\doibase
  10.1103/PhysRevD.90.112005} {\bibfield  {journal} {\bibinfo  {journal} {Phys.
  Rev.}\ }\textbf {\bibinfo {volume} {D90}},\ \bibinfo {pages} {112005}
  (\bibinfo {year} {2014})},\ \Eprint {http://arxiv.org/abs/1409.5542}
  {arXiv:1409.5542 [hep-ex]} \BibitemShut {NoStop}%
\end{thebibliography}%

\end{document}